\documentclass[english,aps,prc,nofootinbib,superscriptaddress,twocolumn,letter]{revtex4}
\usepackage[latin9]{inputenc}
\usepackage{hyperref}
\hypersetup{
  colorlinks=true,        
  linkcolor=blue,         
  citecolor=cyan,         
}
\usepackage{breakurl}
\usepackage{graphicx}
\usepackage{epsf}
\usepackage{epsfig}
\usepackage{amssymb,amsmath}
\usepackage[usenames]{color}
\usepackage{amssymb}
\usepackage{times}
\usepackage{comment}

\newcommand\snn{\sqrt{s_\text{NN}}}

\allowdisplaybreaks



\begin{document}

\preprint{This line only printed with preprint option}

\title{Directed flow and global polarization in Au+Au collisions across energies covered by the beam energy scan at RHIC
}

\author{Ze-Fang Jiang}
\email{jiangzf@mails.ccnu.edu.cn}
\affiliation{Department of Physics and Electronic-Information Engineering, Hubei Engineering University, Xiaogan, Hubei, 432000, China}
\affiliation{Institute of Particle Physics and Key Laboratory of Quark and Lepton Physics (MOE), Central China Normal University, Wuhan, Hubei, 430079, China}

\author{Xiang-Yu Wu}
\affiliation{Institute of Particle Physics and Key Laboratory of Quark and Lepton Physics (MOE), Central China Normal University, Wuhan, Hubei, 430079, China}

\author{Shanshan Cao}
\email{shanshan.cao@sdu.edu.cn}
\affiliation{Institute of Frontier and Interdisciplinary Science, Shandong University, Qingdao, Shandong, 266237, China}

\author{Ben-Wei Zhang}
\affiliation{Institute of Particle Physics and Key Laboratory of Quark and Lepton Physics (MOE), Central China Normal University, Wuhan, Hubei, 430079, China}
\affiliation{Guangdong Provincial Key Laboratory of Nuclear Science, Institute of Quantum Matter,
South China Normal University, Guangzhou, Guangdong, 510006, China}

\begin{abstract}
We study the directed flow of identified particles in Au+Au collisions at $\snn=7.7$ to 62.4~GeV. The Glauber model is extended to include both a tilted deformation of the QGP fireball with respect to the longitudinal direction and a non-zero longitudinal flow velocity gradient in the initial state. By combining this improved initial condition with a (3+1)-dimensional viscous hydrodynamic model calculation, we obtain a satisfactory description of the transverse momentum spectra and the rapidity dependent directed flow coefficient of different hadron species. Our calculation indicates the sensitivity of the hadron directed flow, especially its splitting between protons and antiprotons, to both the initial geometry and the initial longitudinal flow velocity. Therefore, the combination of directed flow of different hadrons can provide a tight constraint on the initial condition of nuclear matter created in heavy-ion collisions. The initial condition extracted from the directed flow is further tested with the global polarization of $\Lambda$ and $\bar{\Lambda}$ within the same theoretical framework, where we obtain a reasonable description of these hyperon polarization observed at different collision energies at RHIC. 
\end{abstract}
\maketitle
\date{\today}

\section{Introduction}
\label{emsection1}

A new state of strongly coupled nuclear matter, known as the Quark-Gluon Plasma (QGP), is created in relativistic heavy-ion collisions at the BNL Relativistic Heavy-Ion Collider (RHIC) and the CERN Large Hadron Collider (LHC)~\cite{Ollitrault:1992bk,Rischke:1995ir,Sorge:1996pc,Bass:1998vz,Aguiar:2001ac,Shuryak:2003xe,Heinz:2013th,Busza:2018rrf}.
Lattice QCD calculations suggest that the transition from hadronic matter to the QGP is a smooth crossover at zero baryon density~\cite{Aoki:2006we}. Whether one may obtain a first-order phase transition at finite baryon density by creating a compressed baryonic matter (CBM) is still an open question~\cite{Friman:2011zz}.
A first-order transition indicates the existence of a ``softest point" in the equation of state (EoS), which may leave signature in the final state observables such as the transverse collective flow of hadrons~\cite{Rischke:1995pe}.
To search for the first-order phase transition, various CBM experiments have been constructed to investigate the QCD phase diagram at high baryon density, such as the Beam Energy Scan (BES) experiment at RHIC~\cite{Odyniec:2015iaa}, Nuclotron-based Ion Collider fAcility (NICA)~\citep{Kekelidze:2016wkp}, Japan Proton Accelerator Research Complex for Heavy-Ion (JPARC-HI)~\citep{J-PARCHeavy-Ion:2016ikk}, Alternating Gradient Synchrontron (AGS)~\cite{E877:1997zjw,E895:2000maf} at BNL and Facility for Antiproton and Ion Research (FAIR)~\citep{CBM:2016kpk}. Various observables have been proposed to seek signals of the first order phase transition and locate the critical endpoint (CEP) in the QCD phase diagram, such as higher-order cumulants 
of conserved charges~\cite{STAR:2021iop}, collective flow of emitted particles~\citep{STAR:2014clz,STAR:2017okv,STAR:2021ozh,Luo:2020pef,Bzdak:2019pkr}, amplification of the light nuclei multiplicity ratio~\cite{Sun:2020zxy}, and even jet quenching in low energy collisions~\cite{STAR:2017ieb,Wu:2022vbu}.

The first-order Fourier coefficient of the azimuthal distribution of particles, known as the rapidity-odd directed flow ($v_{1}$) ~\cite{Voloshin:1994mz,Bilandzic:2010jr,STAR:2004jwm,STAR:2014clz,STAR:2017okv,STAR:2019clv,ALICE:2019sgg,STAR:2019vcp},
is among the most popular observables in analyzing the QGP properties, considering that they are sensitive to the initial size and geometry of the nuclear matter produced in energetic collisions~\cite{Gyulassy:1981nq,Gustafsson:1984ka,Lisa:2000ip,PHENIX:2003qra,ALICE:2010suc,CMS:2012zex,Ollitrault:1992bk,Voloshin:1994mz,Nara:2016phs,Chatterjee:2017ahy,Singha:2016mna,Zhang:2018wlk,Guo:2017mkf,Parida:2022lmt}.
Within hydrodynamic models, the directed flow observed in heavy-ion experiments can be understood with an expanding fireball from an initial energy density that is asymmetric (tilted or shifted) with respect to the beam axis.
The related phenomenological model calculations provide a reasonable description of the charged particle $v_{1}$ measured in Au+Au, Zr+Zr, Ru+Ru and Pb+Pb collisions~\cite{Bozek:2022svy,Bozek:2011ua,Jiang:2021ajc,Jiang:2021foj,Shen:2020jwv,Ryu:2021lnx}.
However, with zero baryon density, the splitting of $v_1$ between baryon and anti-baryon cannot be explained by the deformed initial energy density distribution alone.
It is a great challenge to quantitatively describe the different directed flow coefficients between protons and antiprotons measured at different collision energies at RHIC-BES~\cite{STAR:2014clz}, NA49~\cite{NA49:2003njx} and E895~\cite{E895:2000maf} using earlier hydrodynamic and transport models~\cite{Chen:2009xc,Steinheimer:2014pfa,Konchakovski:2014gda,Guo:2012qi,Ivanov:2014ioa,Nara:2016phs}.
Recently, this problem is investigated for Au+Au collisions at the RHIC energies (7.7-200~GeV) by assuming that the baryon density is also counterclockwise tilted~\cite{Bozek:2022svy} or with specific baryon deposition~\cite{Parida:2022zse,Parida:2022ppj,Du:2022yok} in the reaction plane with respect to the longitudinal direction. Therefore, it is of great interest to further explore whether this proposal can also be verified at other colliding energies, and whether the corresponding initial geometry of nuclear matter consists with other observables such as the global polarization of $\Lambda$ and $\overline{\Lambda}$ hyperons in heavy-ion collisions.

In this work, we investigate the splitting of directed flow between protons and antiprotons in Au+Au collisions across the BES energies ($\snn= 7.7$ - 62.4~GeV) using the (3+1)-dimensional viscous hydrodynamic model CLVisc~\cite{Pang:2016igs,Pang:2018zzo,Wu:2018cpc,Wu:2021fjf}. The 3-D initial condition of the QGP is developed from our earlier study~\cite{Jiang:2021foj} to further include the tilted deformation of the baryon density distribution~\cite{Bozek:2022svy} and the longitudinal flow velocity gradient of the QGP beyond the Bjorken approximation~\cite{Shen:2020jwv,Ryu:2021lnx}. By combining this improved initial condition and the CLVisc model, we are able to provide a satisfactory description of the transverse momentum ($p_\text{T}$) spectra of identified hadrons ($\pi^+$, $K^+$, $p$ and $\bar{p}$) in different centrality classes of Au-Au collisions at the BES energies.
We further show that a simultaneous description of $v_1$ of pions, protons and antiprotons relies on the initial geometry of both the medium energy density and the baryon number density, and the initial longitudinal flow velocity profile. In the end, the medium geometry and longitudinal flow constrained from the rapidity ($y$) dependence of $v_1$ is further tested by the global polarization of hyperons, from which the correlation between directed flow and global polarization can be inferred.

The rest of this paper is organized as follows.
In Sec.~\ref{v1section2}, we will present our modified Glauber model for initializing the QGP, and the CLVisc hydrodynamic model simulation of its further evolution.
In Sec.~\ref{v1section3}, we will calculate the transverse momentum spectra, directed flow of identified particles and global polarization of $\Lambda$ and $\overline{\Lambda}$ hyperons measured in relativistic heavy-ion collisions at the BES energies, and investigate their dependence on the initial geometry and longitudinal flow of the QGP. In the end, we will summarize and discuss necessary future developments in Sec.~\ref{v1section4}.

\section{Model framework}
\label{v1section2}

\subsection{Initial condition}
\label{v1subsect2}

We use a modified Glauber model to generate the initial condition of the QGP fireball, which is tilted in the reaction plane of nuclear collisions~\cite{Bozek:2010bi,Jiang:2021foj,Jiang:2021ajc}.
The nuclear thickness function of an incoming nucleus is obtained using the Woods-Saxon (WS) distribution of nucleons as
\begin{equation}
\begin{aligned}
T(x,y)=\int_{-\infty}^{\infty}dz\frac{n_{0}}{1+\exp\left(\frac{r-R_{0}}{d_{0}}\right)},
\label{eq:thicknessf}
\end{aligned}
\end{equation}
where $n_0$ is the average nucleon density, $r=\sqrt{x^{2}+y^{2}+z^{2}}$ is the radial position with $x,~y,~z$ being the space coordinates, $d_{0}$ is the surface diffusiveness parameter, and $R_{0}$ is the radius parameter of the nucleus. For Au+Au collision systems at the BES energies (7.7 - 62.4~GeV), the parameters are listed in Table~\ref{t:parameters}.

\begin{table}[!h]
\begin{center}
\begin{tabular}{| c| c |c| c| c| c | c |}
\hline
\hline
Nucleus            & $n_{0}$ [1/fm$^{3}$]    & $R_{0}$~[fm]    & $d_{0}$ [fm]    \\ \hline
$^{197}_{79}$Au    & 0.17                    & 6.38        & 0.535      \\
\hline
\hline
\end{tabular}
\caption{\label{t:parameters} Parameters of the Woods-Saxon distribution for the Au nucleus~\cite{Loizides:2017ack,STAR:2021mii}.}
\end{center}
\end{table}

For two nuclei moving along the beam direction ($\pm \hat{z}$) and colliding with an impact parameter $\mathbf{b}$, their corresponding thickness functions can be expressed as
\begin{equation}
\begin{aligned}
T_{+}(\mathbf{x}_\text{T})=T(\mathbf{x}_\text{T}-\mathbf{b}/2),~~~~T_{-}(\mathbf{x}_\text{T})=T(\mathbf{x}_\text{T}+\mathbf{b}/2),
\label{eq:t+}
\end{aligned}
\end{equation}
where $\mathbf{x}_\text{T}=(x,y)$ is the transverse plane coordinate. According to the Glauber model, the density distributions of participant nucleons from the two nuclei are then
\begin{align}
T_{1}(\mathbf{x}_\text{T})&=T_{+}(\mathbf{x}_\text{T})\left\{1-\left[1-\frac{\sigma_\text{NN} T_{-}(\mathbf{x}_\text{T})}{A}\right]^{A}\right\}  \, ,\\
T_{2}(\mathbf{x}_\text{T})&=T_{-}(\mathbf{x}_\text{T})\left\{1-\left[1-\frac{\sigma_\text{NN} T_{+}(\mathbf{x}_\text{T})}{A}\right]^{A}\right\}  \, ,
\end{align}
in which $A$ is the mass number and $\sigma_\text{NN}$ is the inelastic nucleon-nucleon scattering cross section~\cite{Loizides:2017ack}.

\begin{figure*}[tbh]
\includegraphics[width=0.32\textwidth]{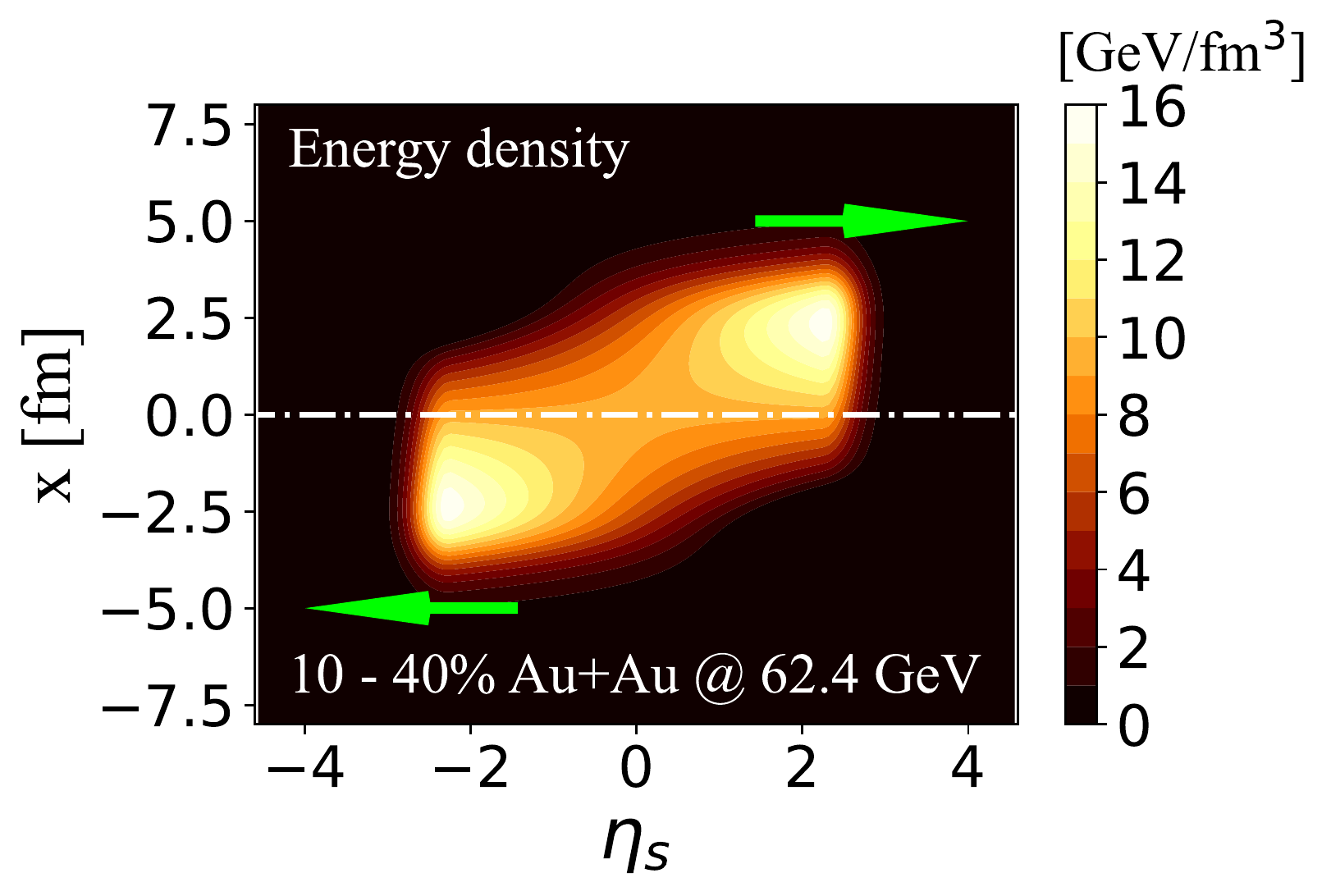}
\includegraphics[width=0.32\textwidth]{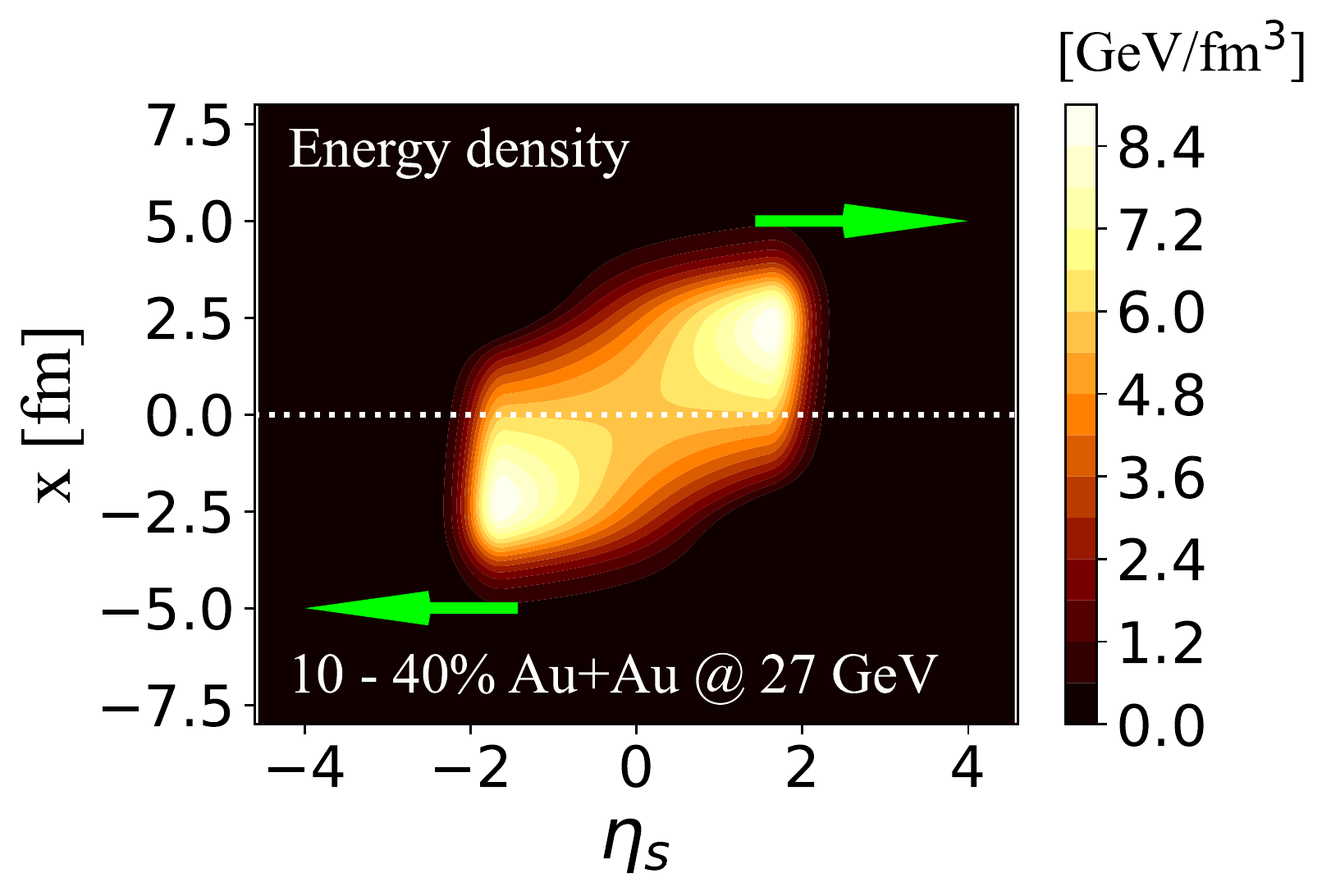}
\includegraphics[width=0.32\textwidth]{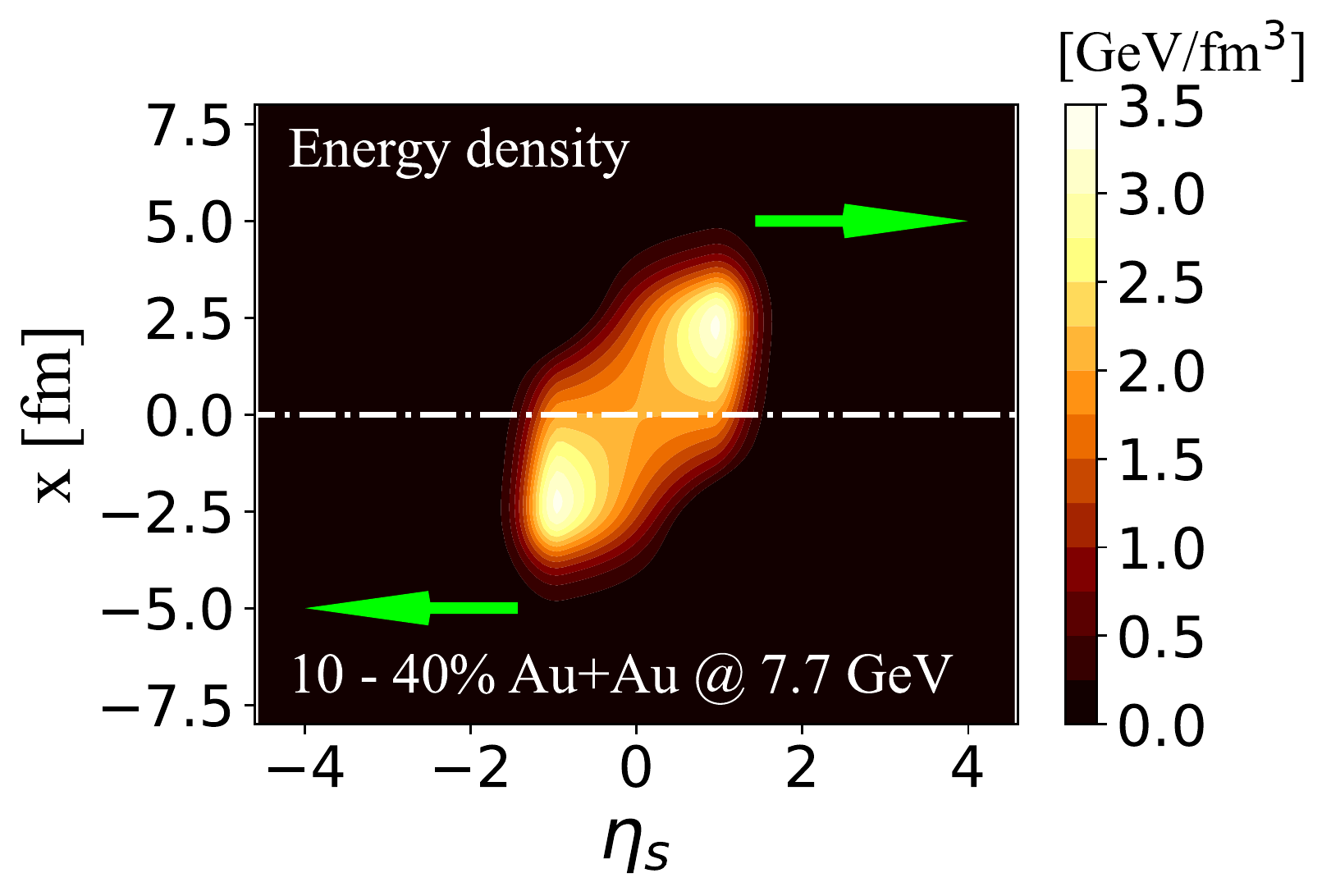}
\vspace{-4pt}
\includegraphics[width=0.32\textwidth]{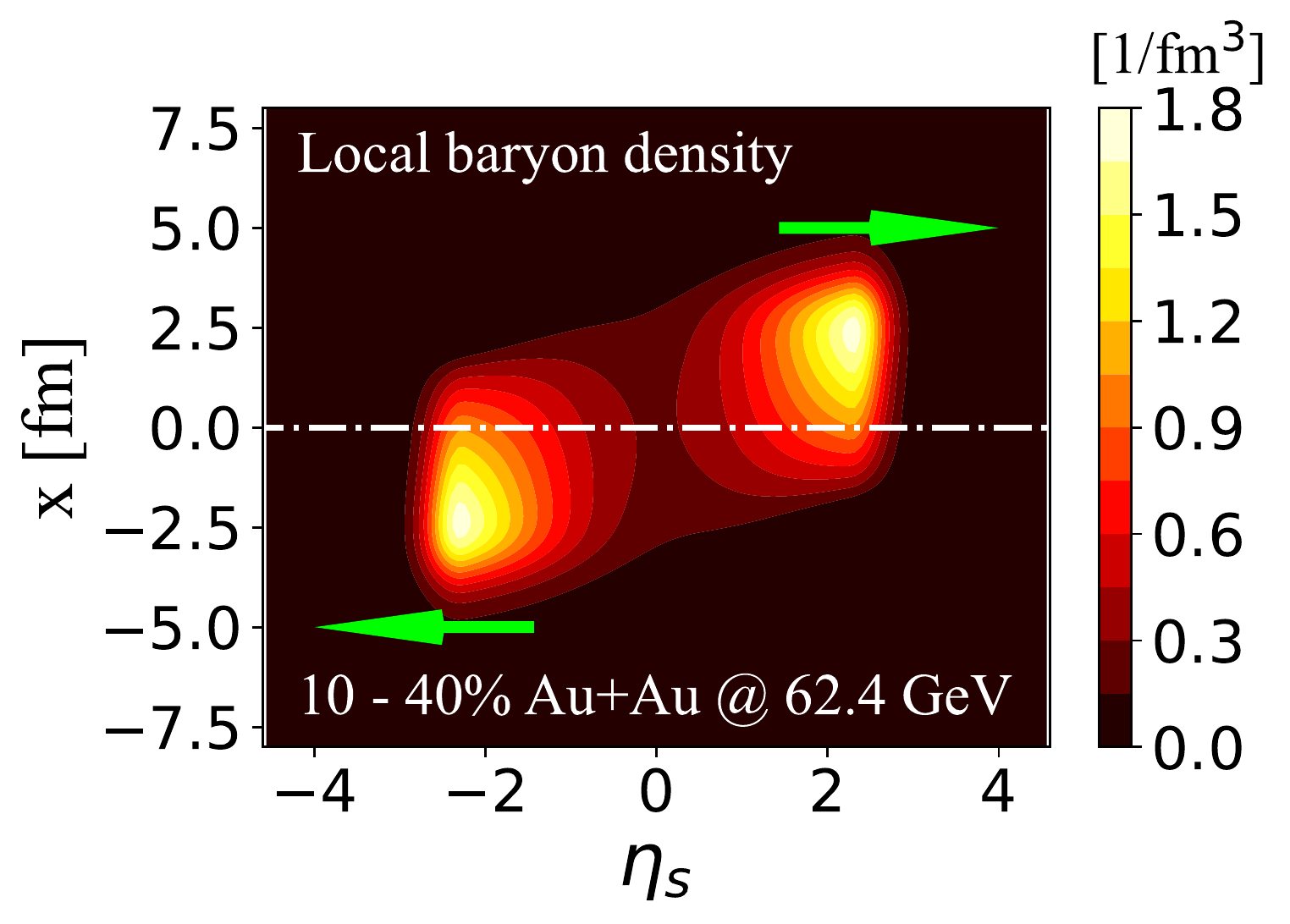}
\includegraphics[width=0.32\textwidth]{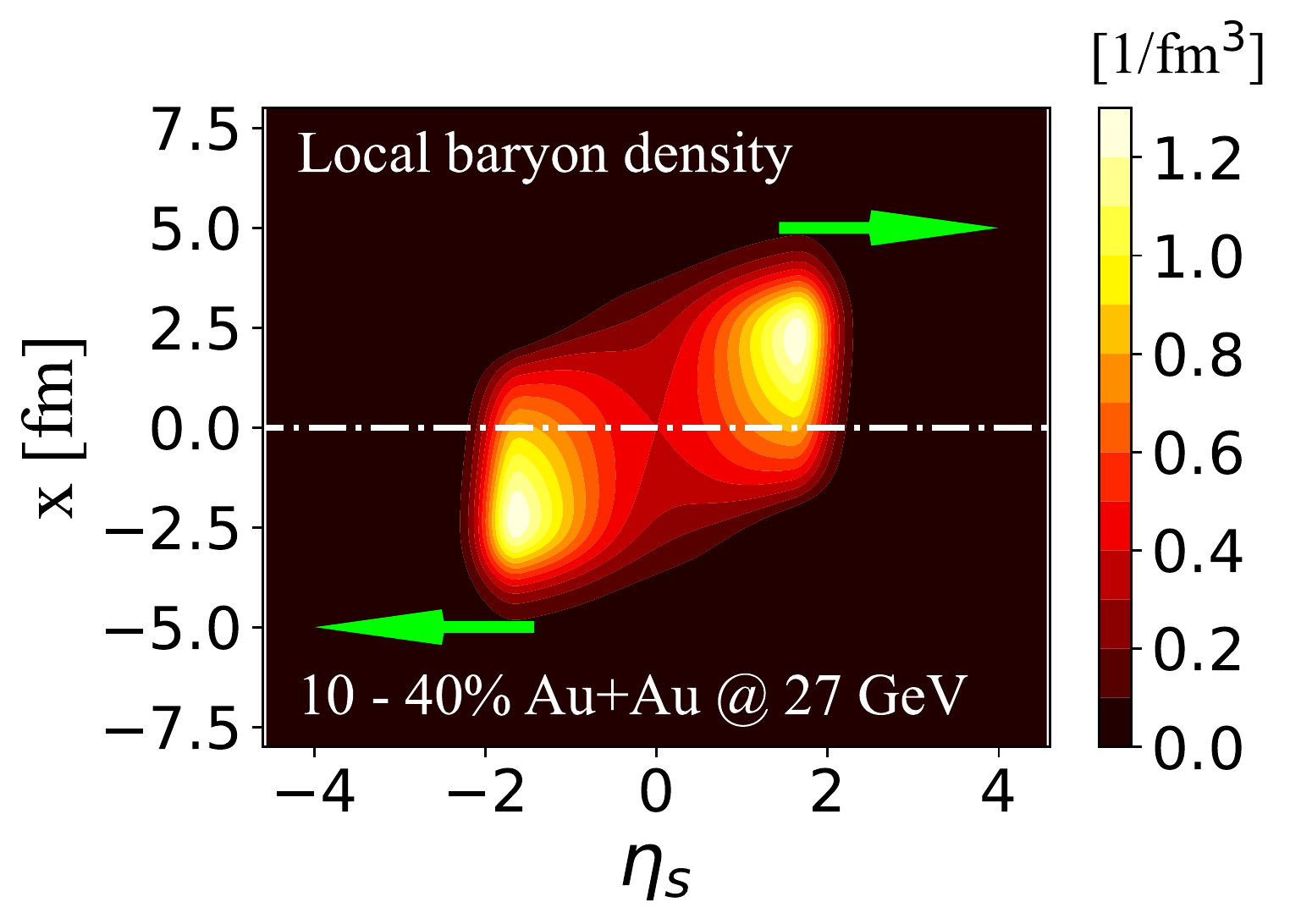}
\includegraphics[width=0.32\textwidth]{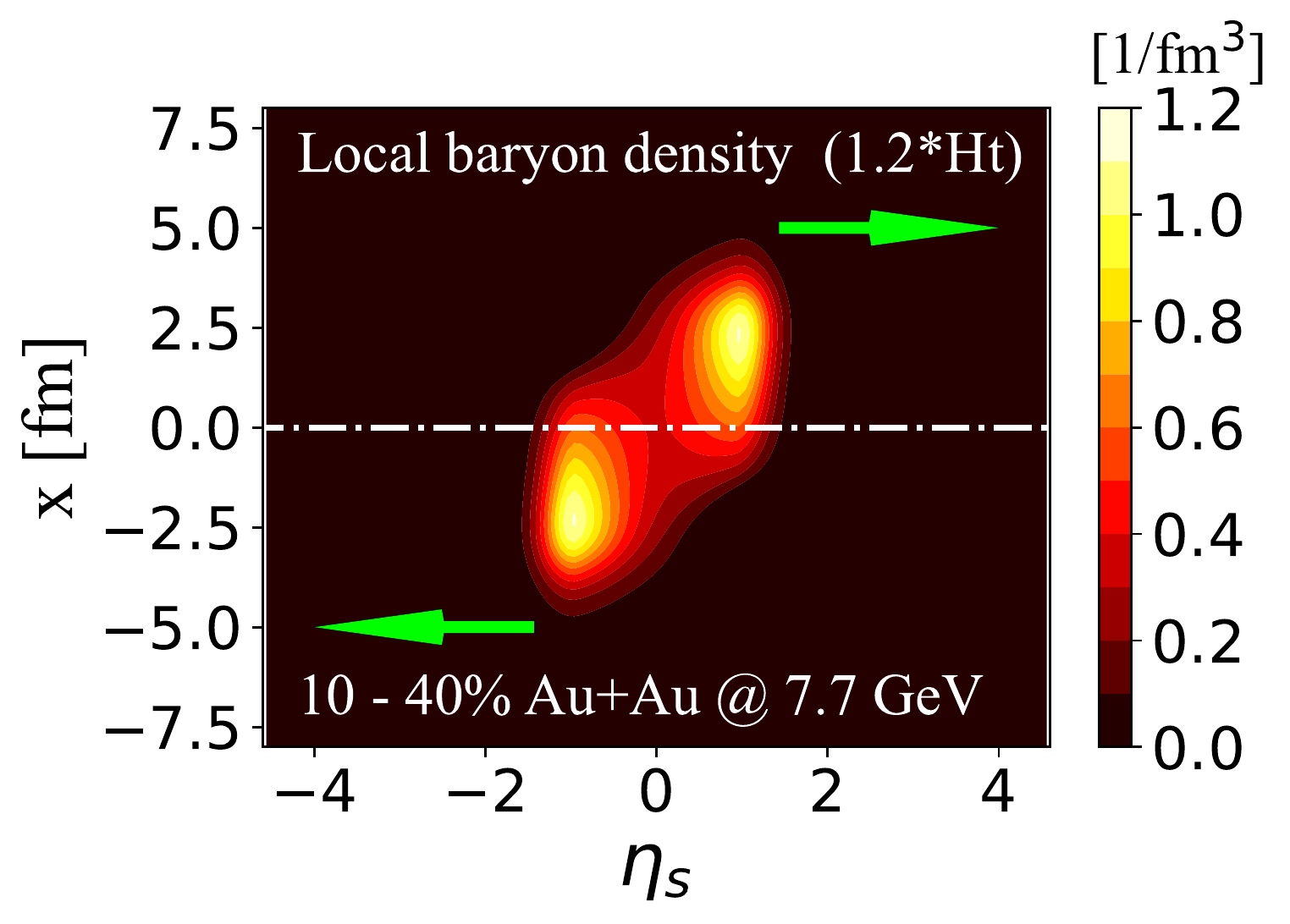}
\caption{(Color online) Distributions of the initial energy density (upper row) and net baryon number density (lower row) on the $\eta_\text{s}$-$x$ plane for 10-40\% Au+Au collisions at $\sqrt{s_\text{NN}}=62.4$, 27 and 7.7~GeV. The arrows (lime color) sketch propagation towards the forward and backward rapidity directions.}
\label{f:auau200ed}
\end{figure*}

Non-central collisions deposit energy into the QGP asymmetrically along the longitudinal direction. As illustrated in Fig.~\ref{f:auau200ed}, a counterclockwise tilt of the medium profile is expected in the reaction plane~\cite{Bozek:2011ua}. 
This deformation can be introduced into the initial condition of the QGP via a rapidity dependent wounded (or participant) nucleon distribution function as~\cite{Jiang:2021foj,Jiang:2021ajc,Jiang:2022uoe} 
\begin{equation}
\begin{aligned}
W_\text{N}(x,y,\eta_\text{s})=&~T_{1}(x,y)+T_{2}(x,y) \\
+&~H_\text{t}[T_{1}(x,y)-T_{2}(x,y)]\tan\left(\frac{\eta_\text{s}}{\eta_\text{t}}\right),
\label{eq:mnccnu}
\end{aligned}
\end{equation}
where the parameter $H_\text{t}$ reflects the overall strength of imbalance between the forward and backward spacetime rapidities ($\eta_\text{s}$), and the function $\tan (\eta_\text{s}/\eta_\text{t})$ models the rapidity dependence of this imbalance. A fixed parameter $\eta_\text{t}=8.0$ will be used in the present study, which provides a reasonable description of the directed flow ($v_1$) of charged particles in our earlier work~\cite{Jiang:2021foj}.


The total weight function $W(x,y,\eta_\text{s})$ combines contributions from wounded nucleons and binary collisions as
\begin{equation}
\begin{aligned}
W(x,y,\eta_\text{s})=\frac{(1-\alpha)W_\text{N}(x,y,\eta_\text{s})+\alpha n_\text{BC}(x,y)}{\left[(1-\alpha)W_\text{N}(0,0,0)+\alpha n_\text{BC}(0,0)\right]|_{\mathbf{b}=0}},
\label{eq:wneta}
\end{aligned}
\end{equation}
where $n_\text{BC}(x,y)=\sigma_\text{NN}T_{+}(x,y)T_{-}(x,y)$ is the number of binary (hard) collisions, and $\alpha=0.05$ is the collision hardness parameter determined by the centrality (or $\mathbf{b}$) dependence of the soft hadron yield~\cite{Pang:2018zzo,Loizides:2017ack}.

The energy density $\varepsilon(x,y,\eta_\text{s})$ and the local baryon density at the initial time then read~\cite{Wu:2021fjf,Pang:2018zzo}
\begin{align}
\label{eq:eps} \varepsilon(x,y,\eta_\text{s})&=K \cdot W(x,y,\eta_\text{s}) \cdot H(\eta_\text{s}) \, ,\\
\label{eq:nb} n(x,y,\eta_\text{s})&=\frac{1}{N}\cdot W(x,y,\eta_\text{s}) \cdot H(\eta_\text{s}) \cdot H_{B}(\eta_\text{s})  \, ,
\end{align}
in which the overall factor $K$ is determined by the multiplicity distribution ($dN_{\textrm{ch}}/d\eta$ or $dN_{\textrm{ch}}/dy$) of soft hadrons, $N$ is a normalization factor constrained by the number of participant nucleons $N_\text{part}$.

In Eqs.~(\ref{eq:eps}) and~(\ref{eq:nb}), a function
\begin{equation}
\begin{aligned}
H(\eta_\text{s})=\exp\left[-\frac{(|\eta_\text{s}|-\eta_\text{w})^{2}}{2\sigma^{2}_{\eta}}\theta(|\eta_\text{s}|-\eta_\text{w}) \right]
\label{eq:heta}
\end{aligned}
\end{equation}
is introduced to describe the plateau structure of the longitudinal distribution of emitted hadrons, in which $\eta_\text{w}$ controls the width of the central rapidity plateau and $\sigma_{\eta}$ determines the width (speed) of the Gaussian decay outside the plateau region~\cite{Pang:2018zzo}. Following the recent study Ref.~\cite{Bozek:2022svy}, the longitudinal dependence of the baryon density is also introduced into the initial condition via
\begin{equation}
\begin{aligned}
H_{B}(\eta_\text{s})=\exp\left[-\frac{(\eta_\text{s}-\eta_{n})^{2}}{2\sigma^{2}_{n}}\right]+\exp\left[-\frac{(\eta_\text{s}+\eta_{n})^{2}}{2\sigma^{2}_{n}}\right],
\label{eq:heta}
\end{aligned}
\end{equation}
where parameters $\eta_{n}$ and $\sigma_{n}$ are calibrated by the $p_\text{T}$ spectra of protons and antiprotons.
This phenomenological ansatz~\cite{Bozek:2022svy} can be qualitatively justified in the string models of the initial state~\cite{Wu:2021fjf,Shen:2017bsr,Bialas:2004kt}, considering that the titled energy and baryon density profiles originate from strings that connect valence and sea quarks inside nuclei.

Since we aim at understanding the directed flow of soft hadrons and the hyperon polarization within the same framework, the latter of which is sensitive to the gradient of fluid velocity in the longitudinal direction~\cite{Li:2022pyw}, it is necessary to extend the initialization of fluid velocity beyond the Bjorken approximation. Following Refs.~\cite{Shen:2020jwv, Ryu:2021lnx, Alzhrani:2022dpi}, the longitudinal fluid velocity at the initial proper time $\tau_{0}$ can be given by $v_{\eta_\text{s}}=T^{\tau\eta_\text{s}}/(T^{\tau\tau}+P)$ with $P$ being the pressure, and the energy-momentum tensor components read
\begin{align}
\label{eq:Ttautau}
T^{\tau\tau}&=\varepsilon(x,y,\eta_\text{s})\cosh(y_\text{L}) \, ,\\
T^{\tau\eta_\text{s}}&=\frac{1}{\tau_{0}}\varepsilon(x,y,\eta_\text{s})\sinh(y_\text{L})  \, ,
\end{align}
where the rapidity variable is parameterized with
\begin{equation}
\begin{aligned}
y_\text{L} \equiv f_{v} y_{\textrm{CM}}.
\label{eq:yl}
\end{aligned}
\end{equation}
Here, the center of mass rapidity $y_{\textrm{CM}}$ is related to the participant thickness function imbalance as 
\begin{equation}
\begin{aligned}
y_{\textrm{CM}}=\textrm{arctanh} \left[\frac{T_{1}-T_{2}}{T_{1}+T_{2}} \tanh (y_{\textrm{beam}})\right],
\label{eq:ycm}
\end{aligned}
\end{equation}
in which $y_{\textrm{beam}}\equiv\textrm{arccosh}[\sqrt{s_{\textrm{NN}}}/(2m_{\textrm{N}})]$ is the beam rapidity with $m_{\textrm{N}}$ being the nucleon mass; and $f_{v} \in [0, 1]$ models the fractional longitudinal momentum attributed to the corresponding flow velocity.  This $f_{v}$ parameter allows us to vary the magnitude of the longitudinal flow velocity, which further affects the slope of the directed flow with respect to rapidity ($dv_{1}/dy$) and the global polarization of hyperons. For $f_{v}=0$, one has $y_\text{L}=0$, and the velocity field is reduced to the Bjorken flow scenario~\cite{Shen:2020jwv}. The initial fluid velocity in transverse directions are still set as 0 via $T^{\tau x} = T^{\tau y} = 0$, considering that they have little impact on the observables we investigate in this work.

In Table~\ref{table:parameters}, we summarize the parameters used for initializing the QGP produced at the BES energies. The first four parameters ($K$, $\tau_0$, $\sigma_\eta$ and $\eta_\text{w}$) are adjusted according to rapidity dependence of the charged particle yields ($dN_\text{ch}/dy$) in the most central collisions at each colliding energy, and the next two parameters ($\sigma_{n}$ and $\eta_{n}$) are from the $p_\text{T}$ spectra of protons and antiprotons. The last two parameters ($f_v$  and $H_\text{t}$) are determined by the directed flow of hadrons. Note that since we include both the geometric tilt and the longitudinal velocity in the initial QGP profile, values extracted for $f_v$ in this work can be different from those in Ref.~\cite{Ryu:2021lnx} where only the latter effect is taken into account. 

\begin{table}[h]
\centering
\vline
\begin{tabular}{c|c|c|c|c|c|c|c|c|}
\hline
$\sqrt{s_\text{NN}}$ [GeV]& $K$ & $\tau_0$ [fm] & $\sigma_\eta$ [fm] & $\eta_\text{w}$ & $\sigma_{n}$ &$\eta_{n}$ & $f_v$  & $H_\text{t}$\\ \hline
62.4 & 11.8  & 1.0 & 0.3 & 2.25  & 1.34  &2.7  &0.18  &10.0\\ \hline
39   & 8.25  & 1.3 & 0.3 & 1.9   & 1.13  &2.1  &0.22  &13.0\\ \hline
27   & 7.40  & 1.4 & 0.3 & 1.6   & 1.06  &1.8  &0.23  &13.5\\ \hline
19.6 & 5.60  & 1.8 & 0.3 & 1.3   & 0.85  &1.5  &0.24  &15.5\\ \hline
14.5 & 3.90  & 2.2 & 0.3 & 1.15  & 0.81  &1.4  &0.24  &18.0\\ \hline
11.5 & 3.05  & 2.4 & 0.3 & 1.16  & 0.79  &1.22 &0.25  &22.0\\ \hline
7.7  & 2.50  & 2.6 & 0.3 & 0.9   & 0.70  &1.05 &0.26  &28.0\\ \hline
\end{tabular}
\caption{Parameters for the 3-dimensional optical Glauber model of the initial condition of the QGP~\cite{Wu:2021fjf,Shen:2020jwv}.}
\label{table:parameters}
\end{table}

Using these parameterizations, we first present in Fig.~\ref{f:auau200ed} the distributions of the energy density (upper row) and net baryon number density (lower row) at $\tau_0$ on the $\eta_\text{s}$-$x$ plane for 10-40\% Au+Au collisions at three different colliding energies ($\snn = 62.4,~27,~7.7$~GeV).
From the figure, one observes that the energy and baryon densities are not only shifted asymmetrically along the forward and backward rapidity directions, but also tilted counterclockwise in the $\eta_\text{s}$-$x$ plane.
Due to different parametrizations between Eq.~(\ref{eq:eps}) and Eq.~(\ref{eq:nb}), the initial baryon density tends to be shifted towards larger forward and backward rapidity regions than the energy distribution. The asymmetric distribution of baryon density will in the end affect the different abundance between protons and antiprotons at different locations of the QGP fireball~\cite{Bozek:2022svy}. 
Since a stronger drag on the participant nucleons from spectators is expected at lower collisional energies, we obtain a stronger tilt of the density profile and thus a larger $H_\text{t}$ parameter at lower energies. Meanwhile, a larger fractional longitudinal momentum is deposited from the colliding beams into the QGP at lower energies, as reflected by the increasing value of $f_v$ as $\snn$ decreases. At very low energy (for $\snn=11.5$ and 7.7~GeV), we also need to assume the baryon density has a stronger tilt than the energy density by applying $1.2 H_\text{t}$ for the former, in order to improve our phenomenological description of the proton $v_1$. This might result from effects of the phase transition~\cite{Konchakovski:2014gda,Ivanov:2014ioa,Steinheimer:2014pfa,Nara:2016phs}, emission of the spectator matter~\cite{Zhang:2018wlk} and the electromagnetic field~\cite{Rybicki:2013qla} that have not been taken into account in our present work.



\subsection{Hydrodynamic evolution}

Starting with the initial condition constructed in the previous subsection,
we utilize a (3+1)-D viscous hydrodynamic model CLVisc~\cite{Pang:2016igs,Pang:2018zzo,Wu:2018cpc,Wu:2021fjf}
to simulate the subsequent evolution of the QGP medium. The hydrodynamic equations read~\cite{Jiang:2020big,Jiang:2018qxd,Denicol:2012cn,Romatschke:2009im,Romatschke:2017ejr}
\begin{align}
\nabla_{\mu} T^{\mu\nu}&=0 \, ,\\
\nabla_{\mu} J^{\mu}&=0  \, ,
\end{align}
where the energy-momentum tensor $T^{\mu\nu}$ and the net baryon current $J^{\mu}$ are defined as
\begin{align}
T^{\mu\nu} &= \varepsilon U^{\mu}U^{\nu} - P\Delta^{\mu\nu} + \pi^{\mu\nu}\,, \\	
J^{\mu} &= nU^{\mu}+V^{\mu}\,,
\end{align}
with $\varepsilon$, $P$, $n$, $u^{\mu}$, $\pi^{\mu\nu}$, $V^{\mu}$ being the local energy density, pressure, net baryon density, flow velocity field, shear stress tensor and baryon diffusion current respectively.
The projection tensor is given by $\Delta^{\mu\nu} = g^{\mu\nu}-u^{\mu}u^{\nu}$, with $g^{\mu\nu} = \text{diag} (1,-1,-1,-1)$ being the metric tensor. Effects of the bulk viscosity are not included in the present study yet~\citep{Shen:2017bsr, Akamatsu:2018olk, Shen:2020jwv, Denicol:2018wdp, Wu:2021fjf}.

Based on the Israel-Stewart second order hydrodynamic expansion, the dissipative currents $\pi^{\mu\nu}$ and $V^{\mu}$ are expressed as follows~\citep{Denicol:2018wdp}:
\begin{align}
\Delta^{\mu\nu}_{\alpha\beta} (u\cdot \partial) \pi^{\alpha\beta} = &
 -\frac{1}{\tau_{\pi}}\left(\pi^{\mu\nu} - \eta_\text{v}\sigma^{\mu\nu}\right)
- \frac{4}{3}\pi^{\mu\nu}\theta
\nonumber
\\
&
-\frac{5}{7}\pi^{\alpha<\mu}\sigma_{\alpha}^{\nu>}+ \frac{9}{70}\frac{4}{e+P}\pi^{<\mu}_{\alpha}\pi^{\nu>\alpha}\,,
\nonumber
\\
\Delta^{\mu\nu} (u\cdot \partial) V_{\nu}  = &  - \frac{1}{\tau_V}\left(V^{\mu}-\kappa_B\bigtriangledown^{\mu}\frac{\mu_B}{T}\right)-V^{\mu}\theta
\nonumber \\
&-\frac{3}{10}V_{\nu}\sigma^{\mu\nu}\,,
\end{align}
where $\theta = \partial \cdot u$ is the expansion rate, $\sigma^{\mu\nu} = \partial^{<\mu} u^{\nu>}$ is the shear tensor,
$\eta_\text{v}$ and $\kappa_B$ are the shear viscosity and baryon diffusion coefficient.
For an arbitrary tensor $A^{\mu\nu}$, its traceless symmetric part is given by $A^{<\mu\nu>} = \frac{1}{2}[(\Delta^{\mu\alpha}\Delta^{\nu\beta}+\Delta^{\nu\alpha}\Delta^{\mu\beta})-\frac{2}{3}\Delta^{\mu\nu}\Delta^{\alpha\beta}]A_{\alpha \beta}$~\cite{Wu:2021fjf}.

In hydrodynamic simulation, the specific shear viscosity $C_{\eta_\text{v}}$ and the baryon diffusion coefficient $\kappa_B$ are treated as model parameters, which are related to $\eta_\text{v}$ and $C_B$ as:
\begin{align}
C_{\eta_\text{v}} &= \frac{\eta_\text{v} T}{e+P}, \label{eq:C_shear}\\
\kappa_B &= \frac{C_B}{T}n\left[\frac{1}{3} \cot \left(\frac{\mu_B}{T}\right)-\frac{nT}{e+P}\right] \,, \label{eq:CB}
\end{align}
where $\mu_B$ stands for the baryon chemical potential. They connect to the relaxation times as 
\begin{equation}
    \tau_{\pi} = \frac{5C_{\eta_\text{v}}}{T},\;\;\;\tau_V = \frac{C_B}{T}.
\end{equation}
In this work, we set $C_{\eta_\text{v}}=0.08$ and ${C_B}=0.4$ for all collision energies.

The hydrodynamic equations are then solved together with the NEOS-BQS equation of state (EOS)~\cite{Monnai:2019hkn,Monnai:2021kgu}, which is based on the lattice QCD calculation at high temperature and vanishing net baryon density and then extended to finite net baryon density according to the Taylor expansion method~\cite{Monnai:2019hkn,Monnai:2021kgu}.
It connects the QGP and hadron phases with a smooth crossover under the strangeness neutrality ($n_S=0$) and the electric charge density $n_Q = 0.4n_B$ conditions.

\begin{figure*}[htb]
\includegraphics[width=0.45\textwidth]{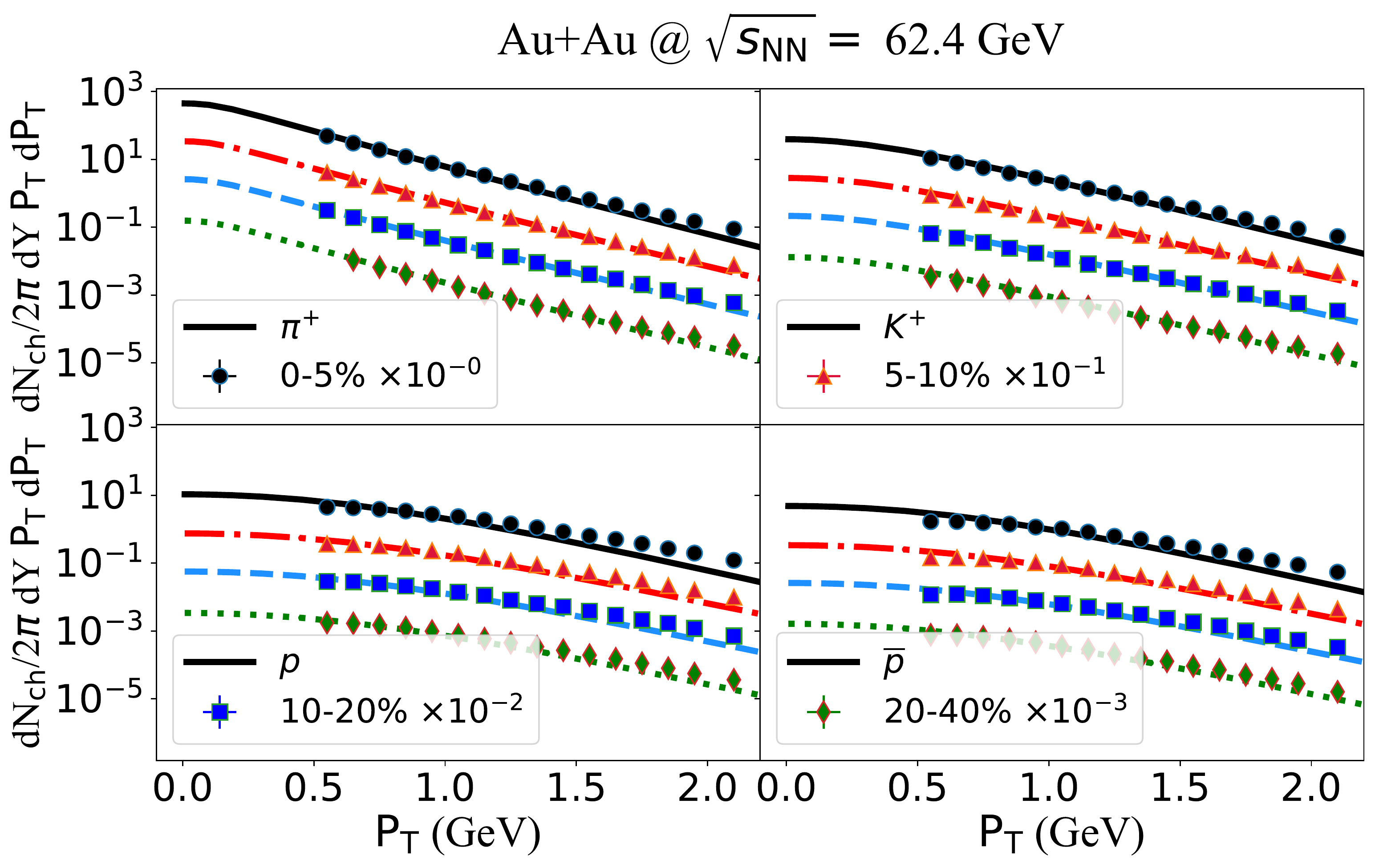}~~~
\includegraphics[width=0.45\textwidth]{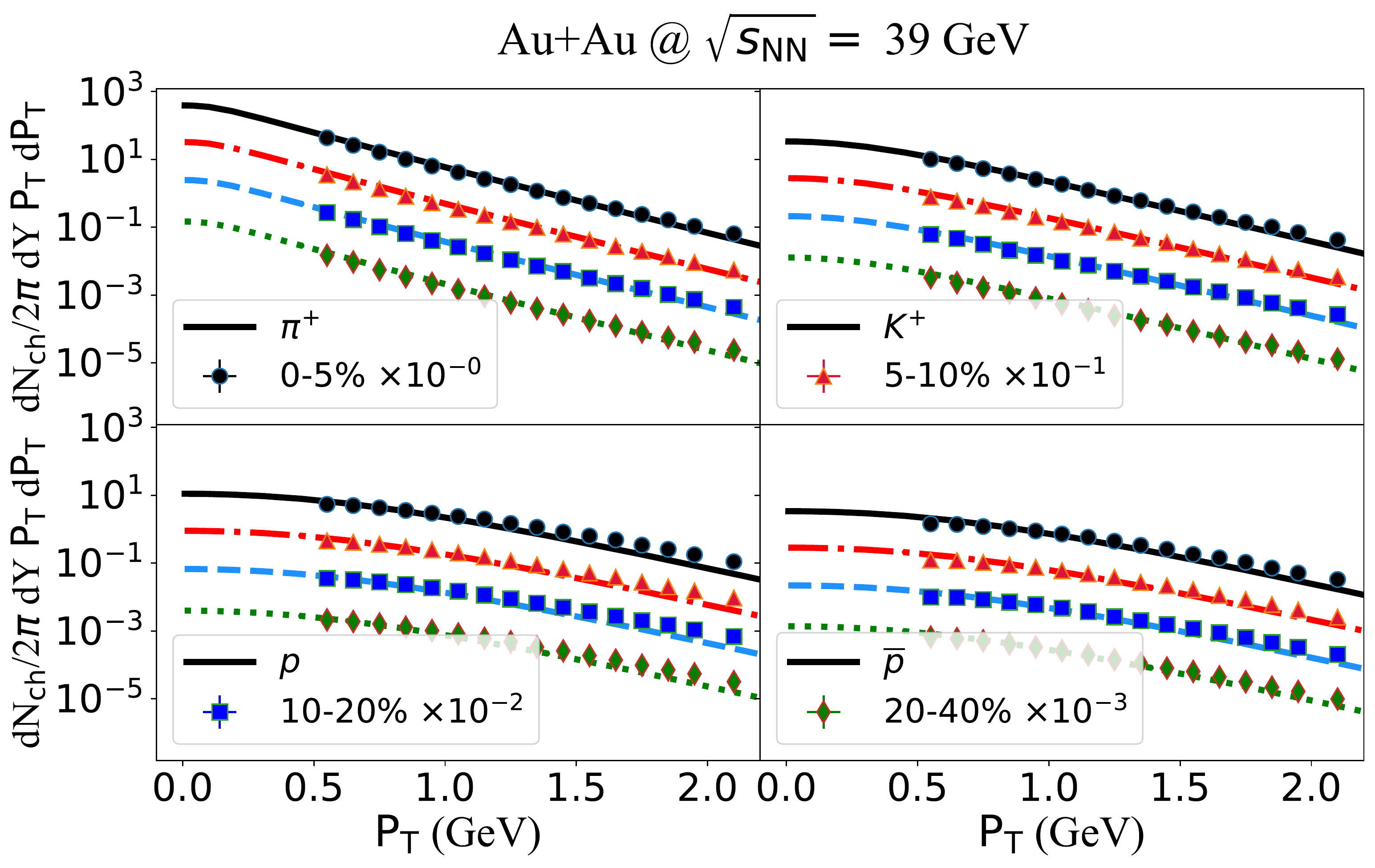}
\vspace{-4pt}
\includegraphics[width=0.45\textwidth]{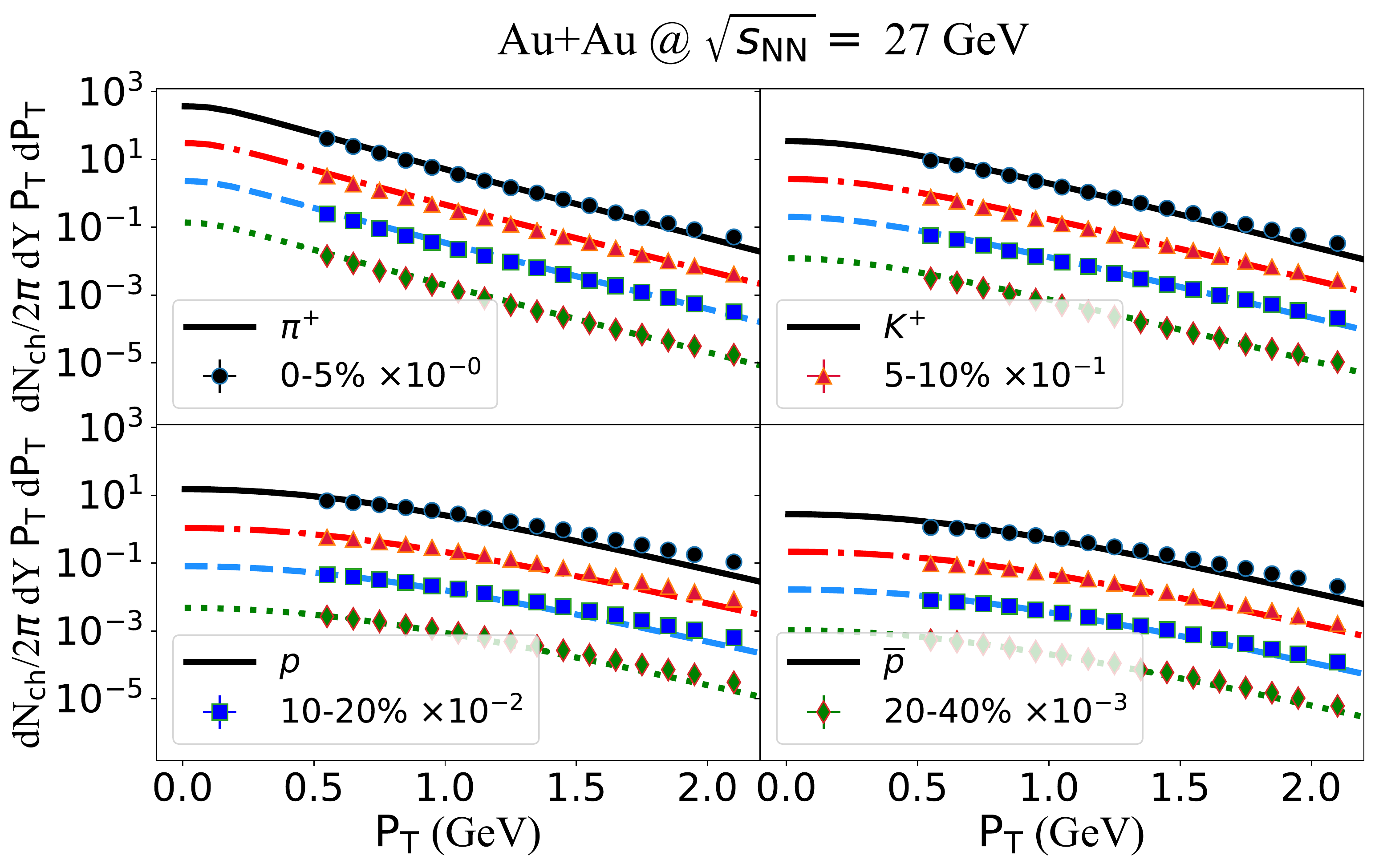}~~~
\includegraphics[width=0.45\textwidth]{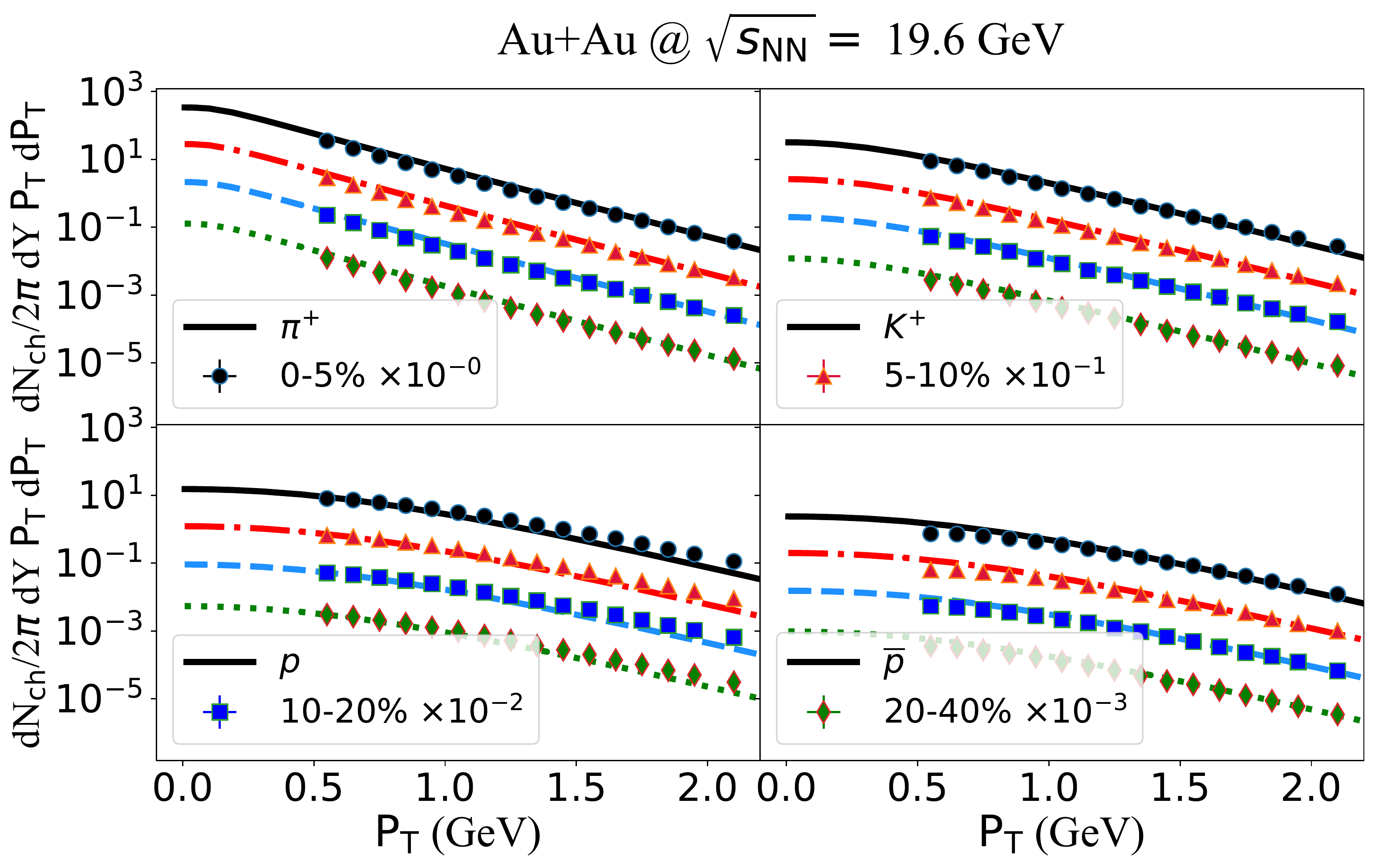}
\vspace{-4pt}
\includegraphics[width=0.45\textwidth]{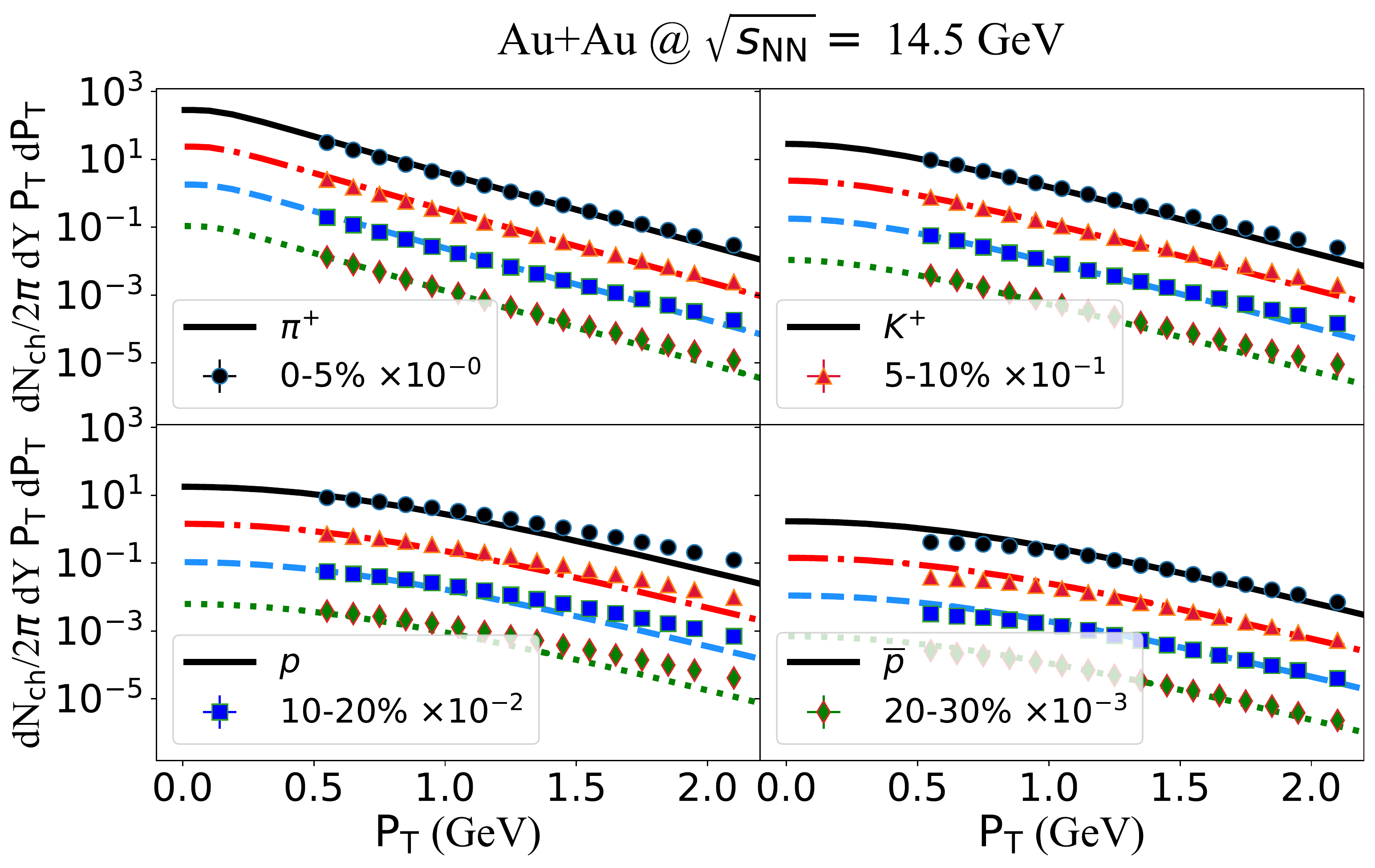}~~~
\includegraphics[width=0.45\textwidth]{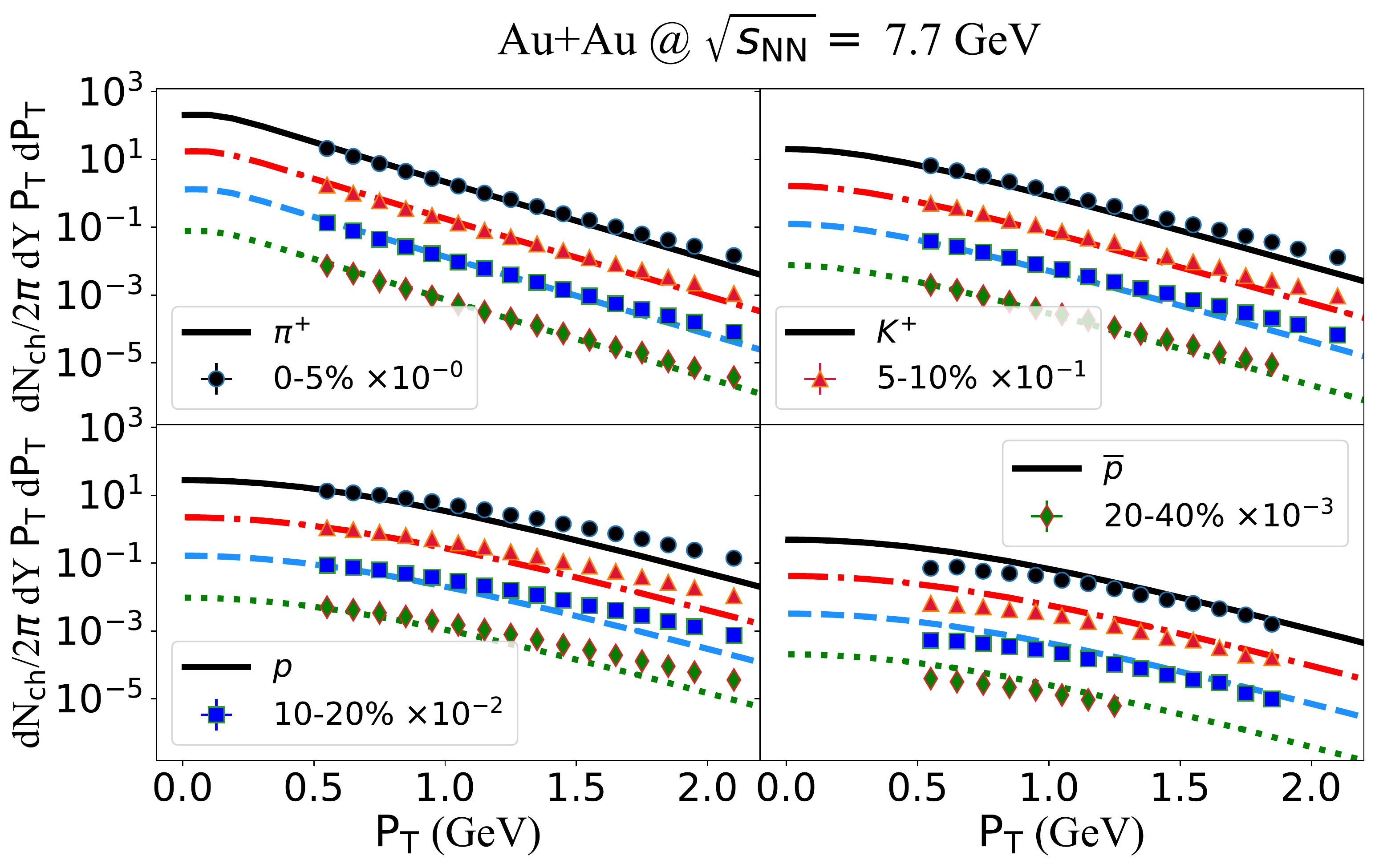}
\caption{(Color online) The transverse momentum spectra of identified charged hadrons ($\pi^+$, $K^+$, $p$ and $\bar{p}$) at mid-rapidity ($|y|<0.25$) in different centrality classes of Au+Au collisions at $\sqrt{s_\text{NN}}= 7.7$, 14.5, 19.6, 27, 39 and 62.4~GeV, compared to the STAR data~\citep{STAR:2017ieb}.}
\label{particle_spectra}
\end{figure*}

\subsection{Particlization}

The isothermal freeze-out condition~\cite{Pang:2018zzo} is applied in our study, with the freeze-out hypersurface  determined by a constant freeze-out energy density ($e_{\text{frz}}$= 0.4~GeV/fm$^3$)~\cite{Wu:2021fjf}. On this hypersurface, the Cooper-Frye formalism is implemented to obtain the momentum distribution of hadrons:
\begin{align}
\frac{dN}{p_\text{T} dp_\text{T} d\phi dy } = \frac{g_i}{(2\pi)^3}\int_{\Sigma} p^{\mu}d\Sigma_{\mu}f_{eq}(1+\delta f_{\pi}+\delta f_{V})\,.
\end{align}
In the above equation, $g_i$ is the spin-color degeneracy factor for identified hadrons, and $d\Sigma_{\mu}$ is the hypersurface element determined by the projection method~\cite{Pang:2018zzo}. The thermal distribution ($f_{\rm eq}$) and its out-of-equilibrium corrections ($\delta f_{\pi}$ and $\delta f_{V}$) are given by
\begin{align}
	f_{\rm eq} &= \frac{1}{\exp \left[(p_{\mu}U^{\mu} - B\mu_B \right)/T_\text{f}] \mp 1} \, ,\\
	\delta f_{\pi}(x,p) &= [1\pm f^{\text{eq}}(x,p)] \frac{p_{\mu}p_{\nu}\pi^{\mu\nu}}{2T^2_f(e+P)}, \\
	\delta f_V(x,p) &= [1\pm f^{\text{eq}}(x,p)]\left(\frac{n_B}{e+P}-\frac{B}{U^{\mu}p_{\mu}}\right)\frac{p^{\mu}V_{\mu}}{\kappa_B/ \tau_V },
\end{align}
where $T_\text{f}$ is the chemical freeze-out temperature, and $B$ represents the baryon number of an identified hadron. 
The out-of-equilibrium corrections above are derived from the Boltzmann equation via the relaxation time approximation~\citep{McNelis:2021acu}. Contributions from resonance decay have been taken into account in this work based on Ref.~\cite{Pang:2018zzo}, although hadronic scatterings after the QGP phase has not been included yet.

\section{Numerical results}
\label{v1section3}

In this section, we present our numerical results on light hadrons in Au+Au collisions at the BES energies
using the (3+1)-D CLVisc hydrodynamics model with finite net baryon density and the tilted initial condition.
We first present the transverse momentum spectra of identified particles $\pi^{+}$, $K^{+}$, $p$ and $\bar{p}$ in Sec.~\ref{sec:3-1}.
Then, we study the rapidity dependence of the directed flow of $\pi^{+}$, $p$ and $\bar{p}$ in Sec.~\ref{sec:3-2}.
The relation between the slope of $v_1$ {\it vs.} $y$ and two competing effects -- the tilt of the fireball ($H_\text{t}$) and the fractional longitudinal momentum transferred into the initial flow velocity ($f_v$) -- are investigated in Sec.~\ref{sec:3-3}.
In the end, we verify the initial condition constrained from the directed flow by reproducing the global polarization of $\Lambda$ and $\bar{\Lambda}$ hyperons in Sec.~\ref{sec:3-4} within the same theoretical framework.

\subsection{Identified particle spectra}
\label{sec:3-1}

We start with validating our model setup by comparing the transverse momentum spectra of the identified light hadrons between our calculation and the STAR data~\citep{STAR:2017ieb} in Fig.~\ref{particle_spectra}.
As discussed in Sec.~\ref{v1subsect2}, the first six model parameters summarized in Tab.~\ref{table:parameters} are adjusted to describe the rapidity dependence of charged particle yields ($dN_\text{ch}/dy$) and the $p_\text{T}$ spectra of proton (antiproton) yield in the most central collisions at each colliding energy. With these parameters, the combination of our initial condition and hydrodynamic evolution is able to provide a reasonable description of the $p_\text{T}$ spectra of different species of identified particles ($\pi^+$, $K^+$, $p$ and $\bar{p}$) across different centrality bins at these colliding energies. 
This implies the success of this model setup in describing the bulk evolution of the nuclear matter produced by heavy-ion collisions at the BES energies. More detailed discussions on $p_\mathrm{T}$ spectra of identified particles, such as their mean $p_\mathrm{T}$, can be found in Ref.~\cite{Wu:2021fjf}, which provides an insight into the radial flow of the QGP medium. This offers a reliable baseline for our further study of the directed flow coefficient and global polarization.

The last two parameters in Tab.~\ref{table:parameters} ($f_{v}$ and $H_\text{t}$) cannot be determined yet, because they control the initial velocity and deformed geometry of the QGP medium, and are insensitive to the integrated particle spectra over the azimuthal angle~\cite{Bozek:2011ua,Jiang:2021ajc,Jiang:2021foj,Bozek:2022svy,Wu:2021fjf,Shen:2020jwv,Ryu:2021lnx}. In other words, varying these two parameters according to the directed flow coefficient later will have little impact on the hadron spectra presented in this subsection.

\begin{figure*}[tbh]
\includegraphics[width=0.32\textwidth]{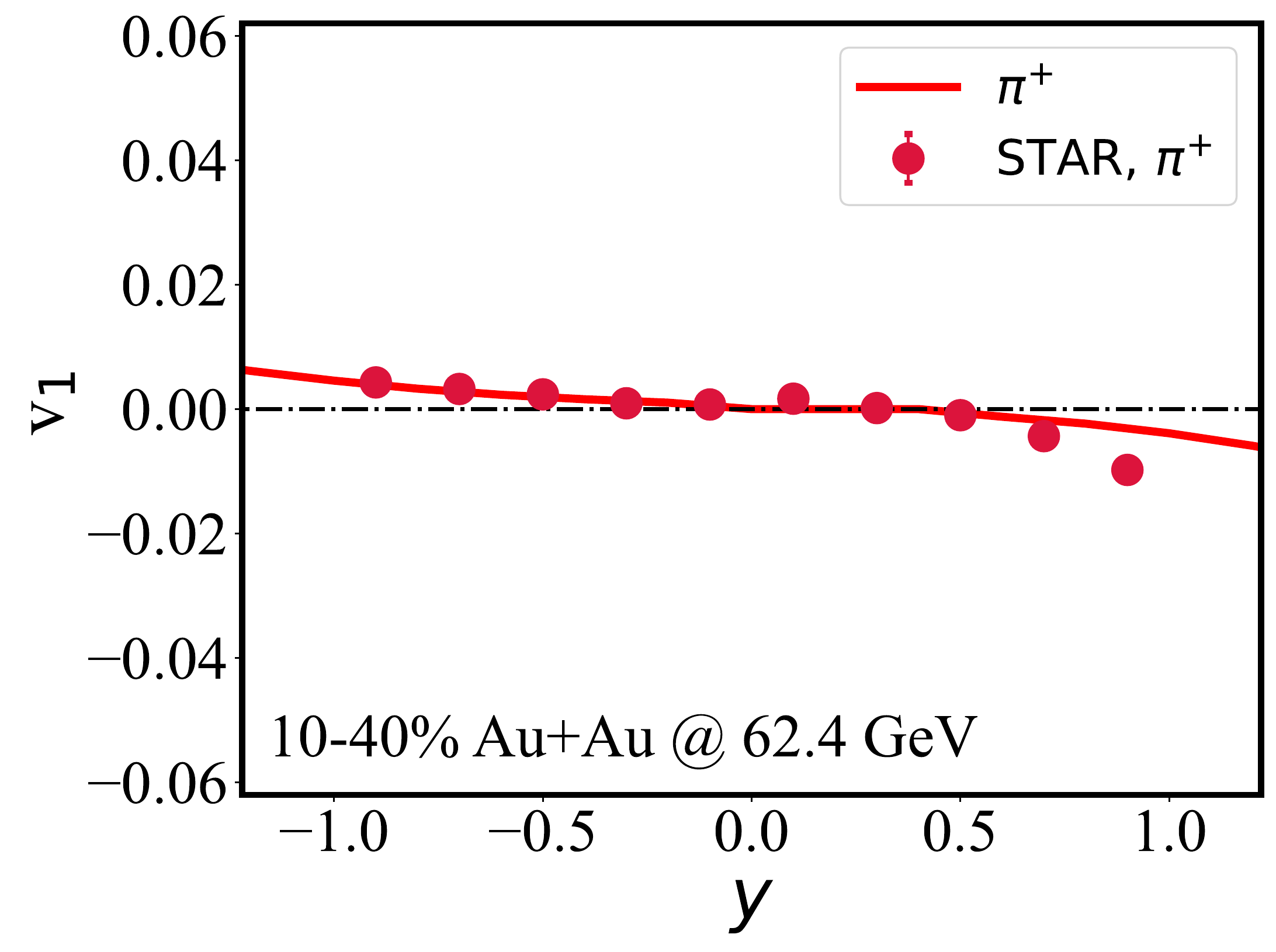}
\includegraphics[width=0.32\textwidth]{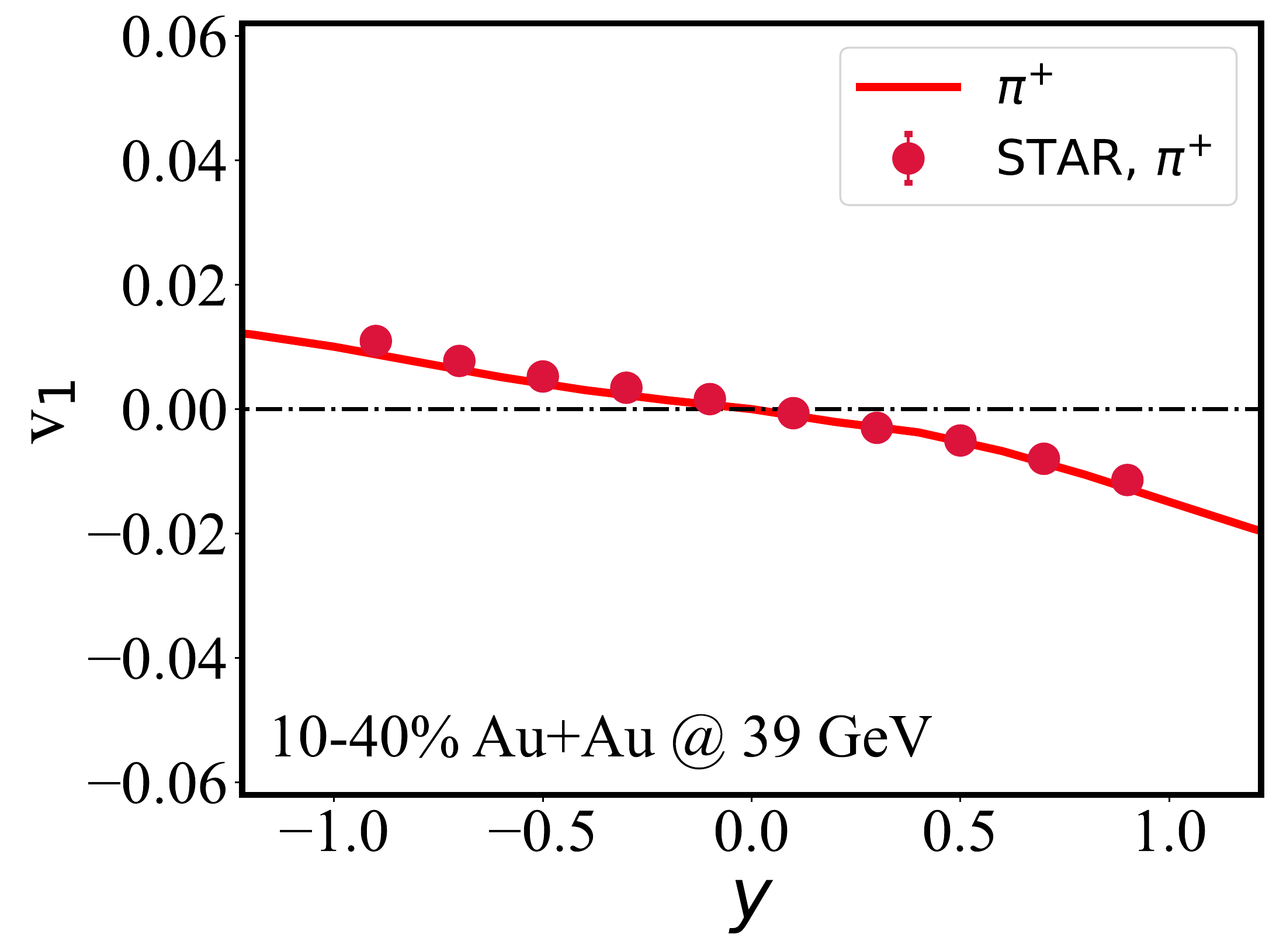}
\includegraphics[width=0.32\textwidth]{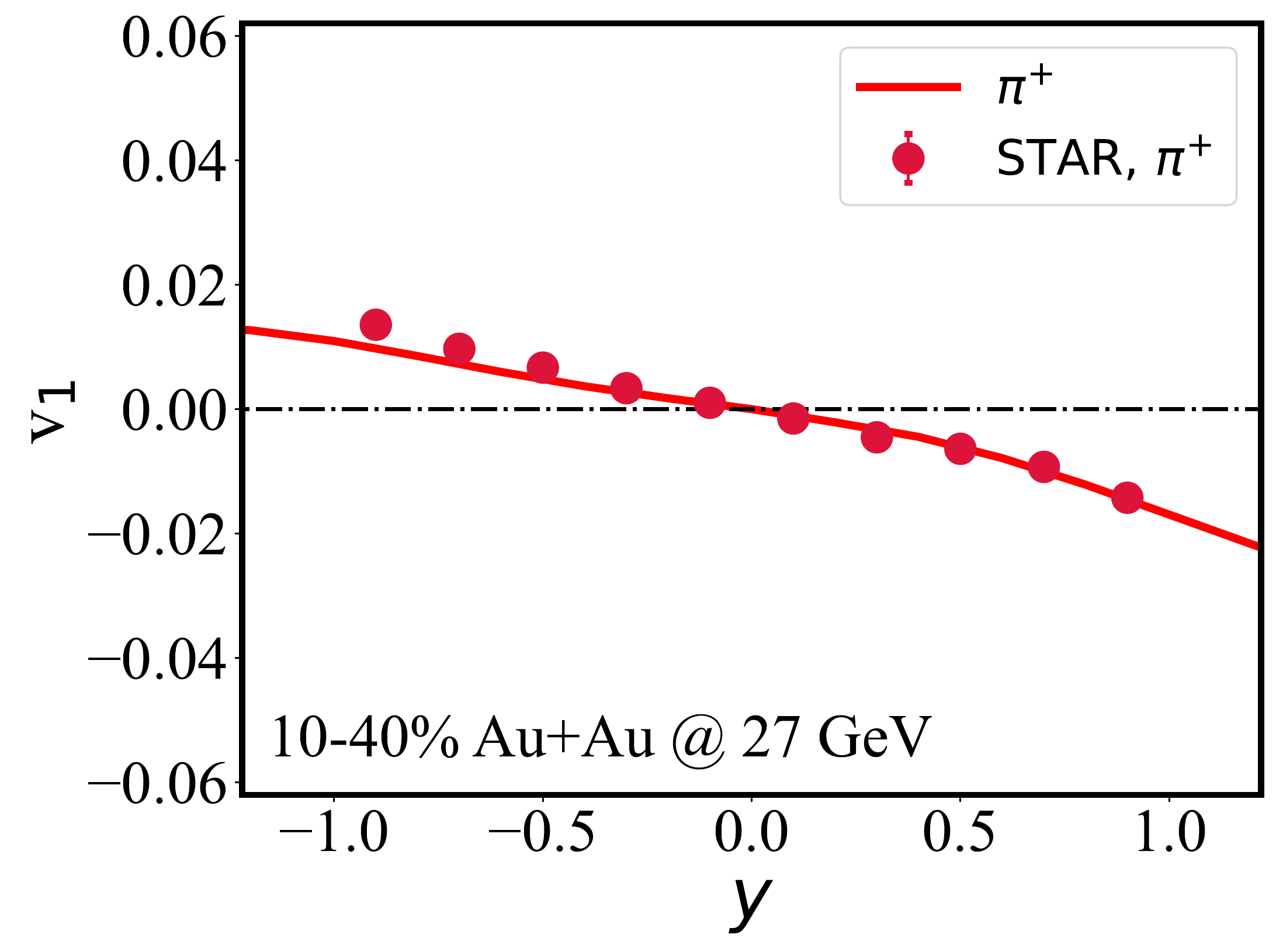}
\vspace{-4pt}
\includegraphics[width=0.32\textwidth]{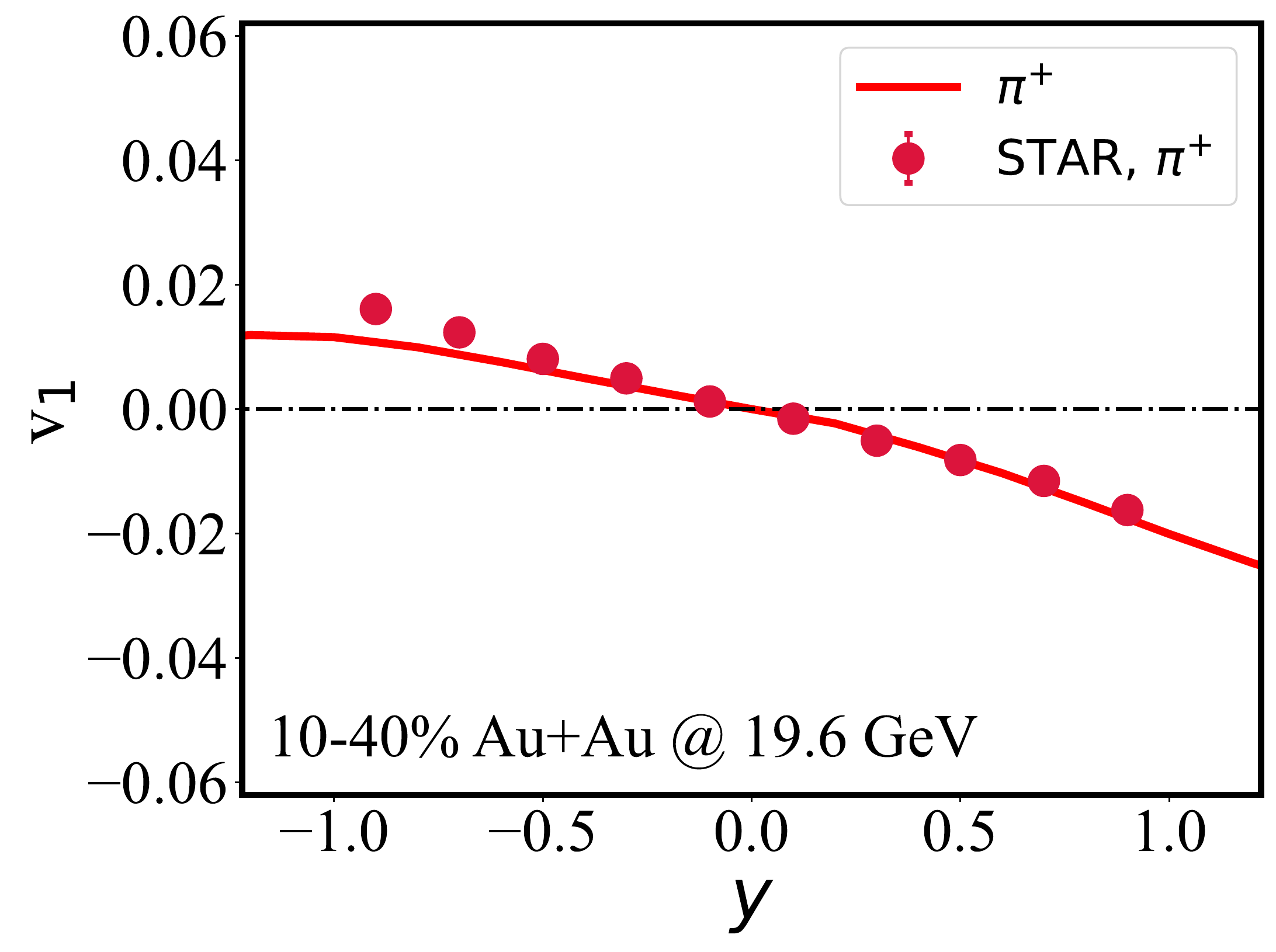}
\includegraphics[width=0.32\textwidth]{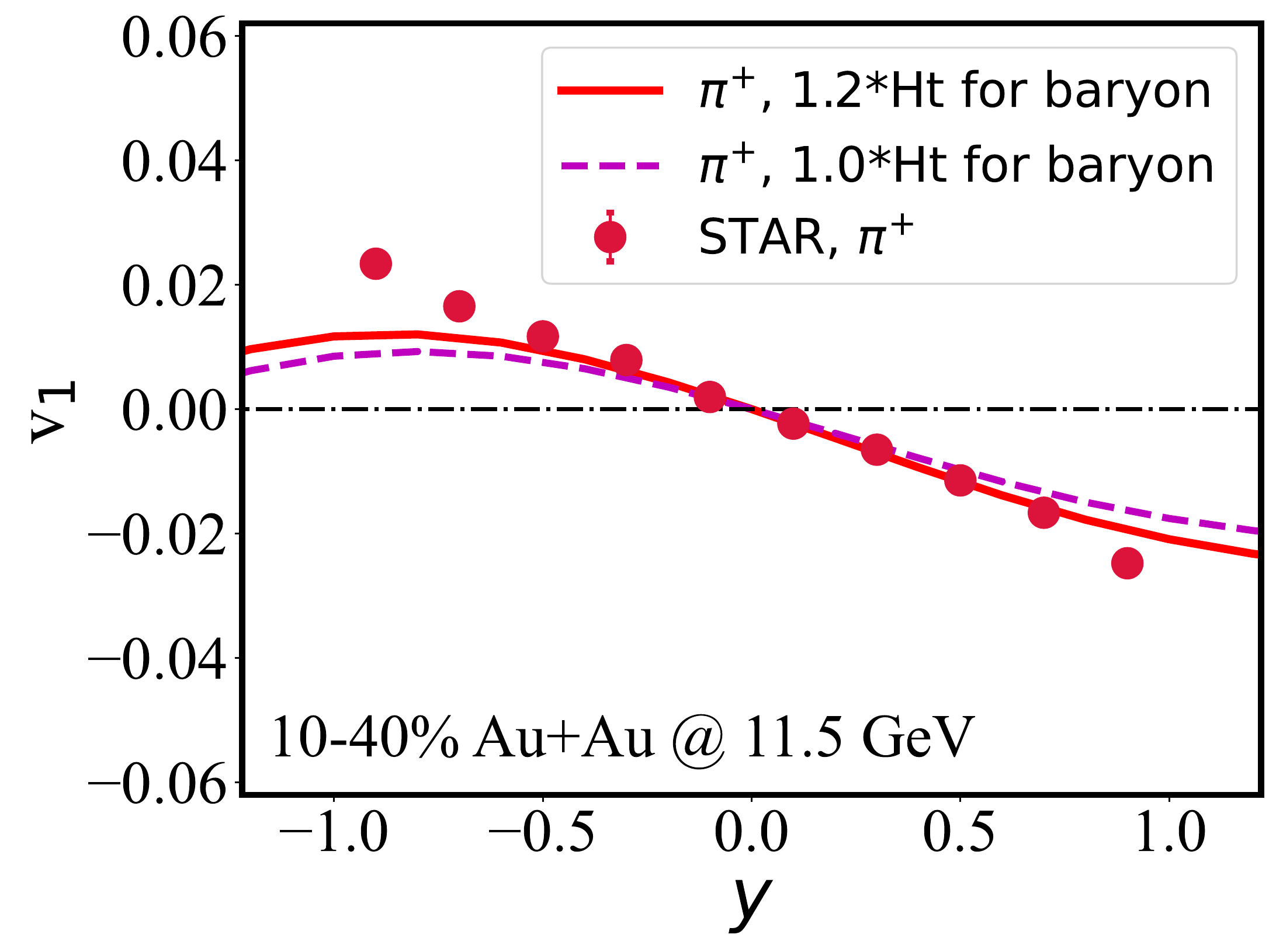}
\includegraphics[width=0.32\textwidth]{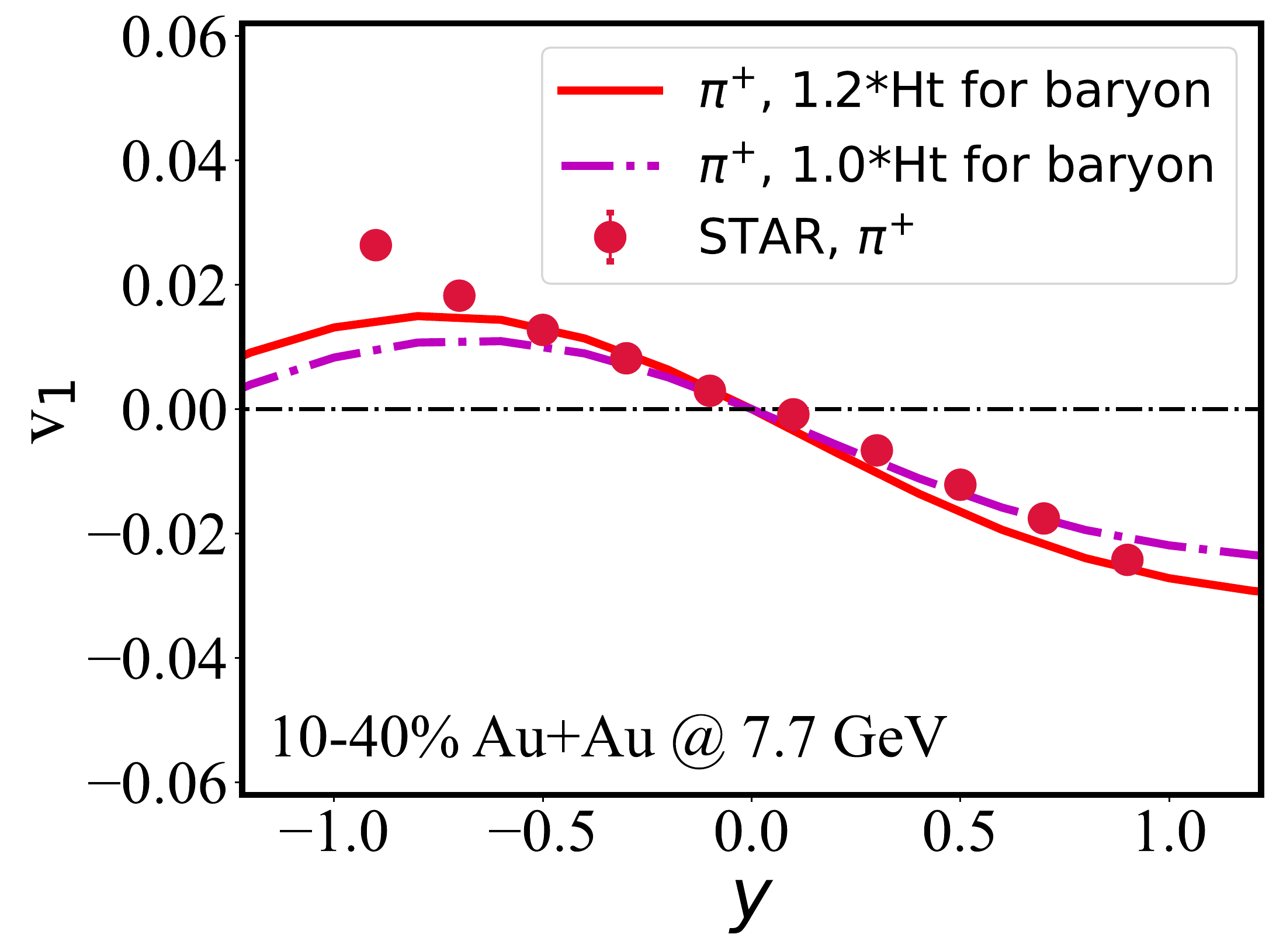}
\caption{(Color online) Rapidity dependence of the directed flow coefficient of $\pi^{+}$ in 10-40\% Au+Au collisions at the BES energies, compared between our model calculation and the STAR data~\cite{STAR:2014clz}.}
\label{f:pionv1}
\end{figure*}

\begin{figure*}[tbh]
\includegraphics[width=0.32\textwidth]{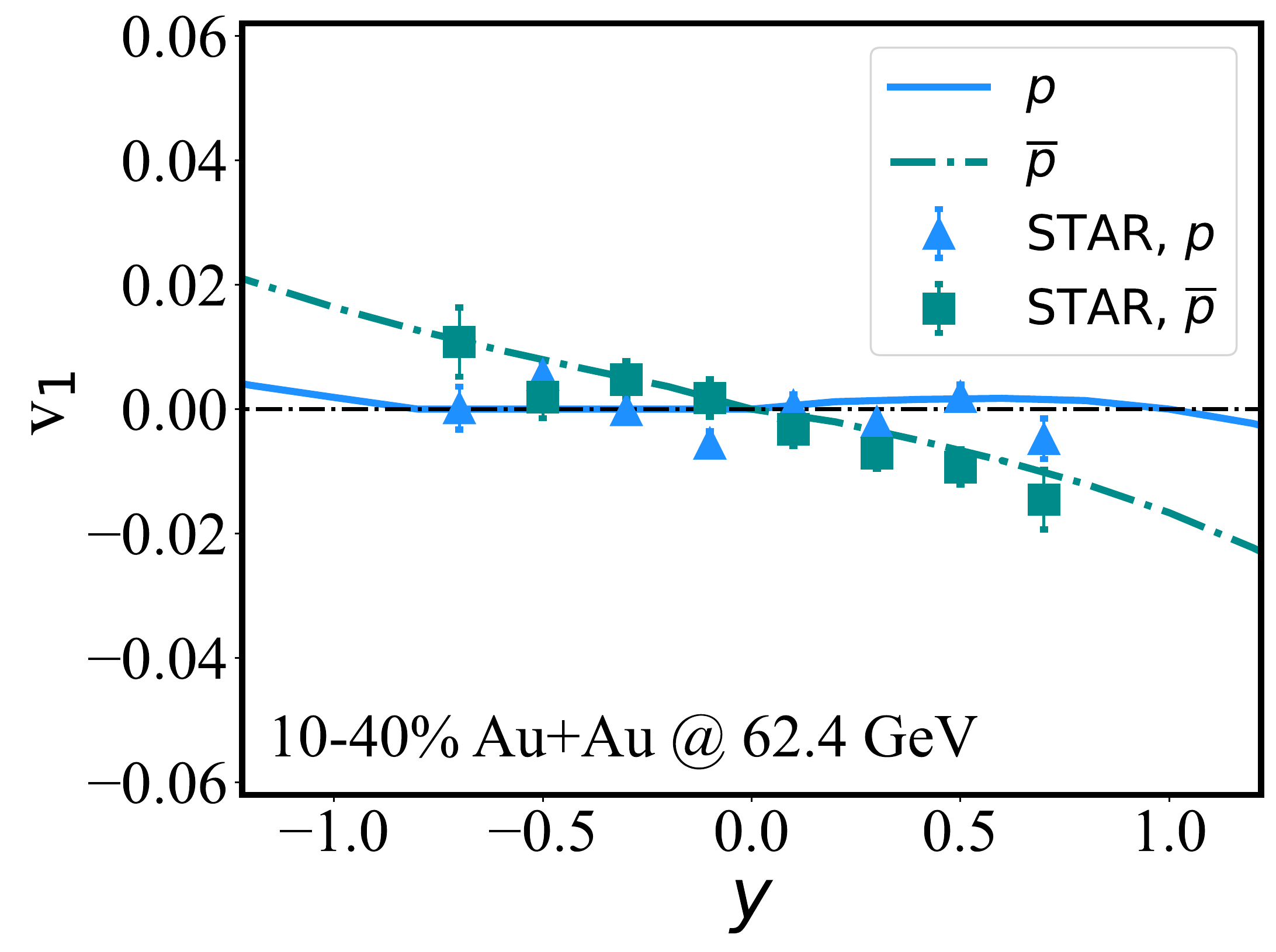}
\includegraphics[width=0.32\textwidth]{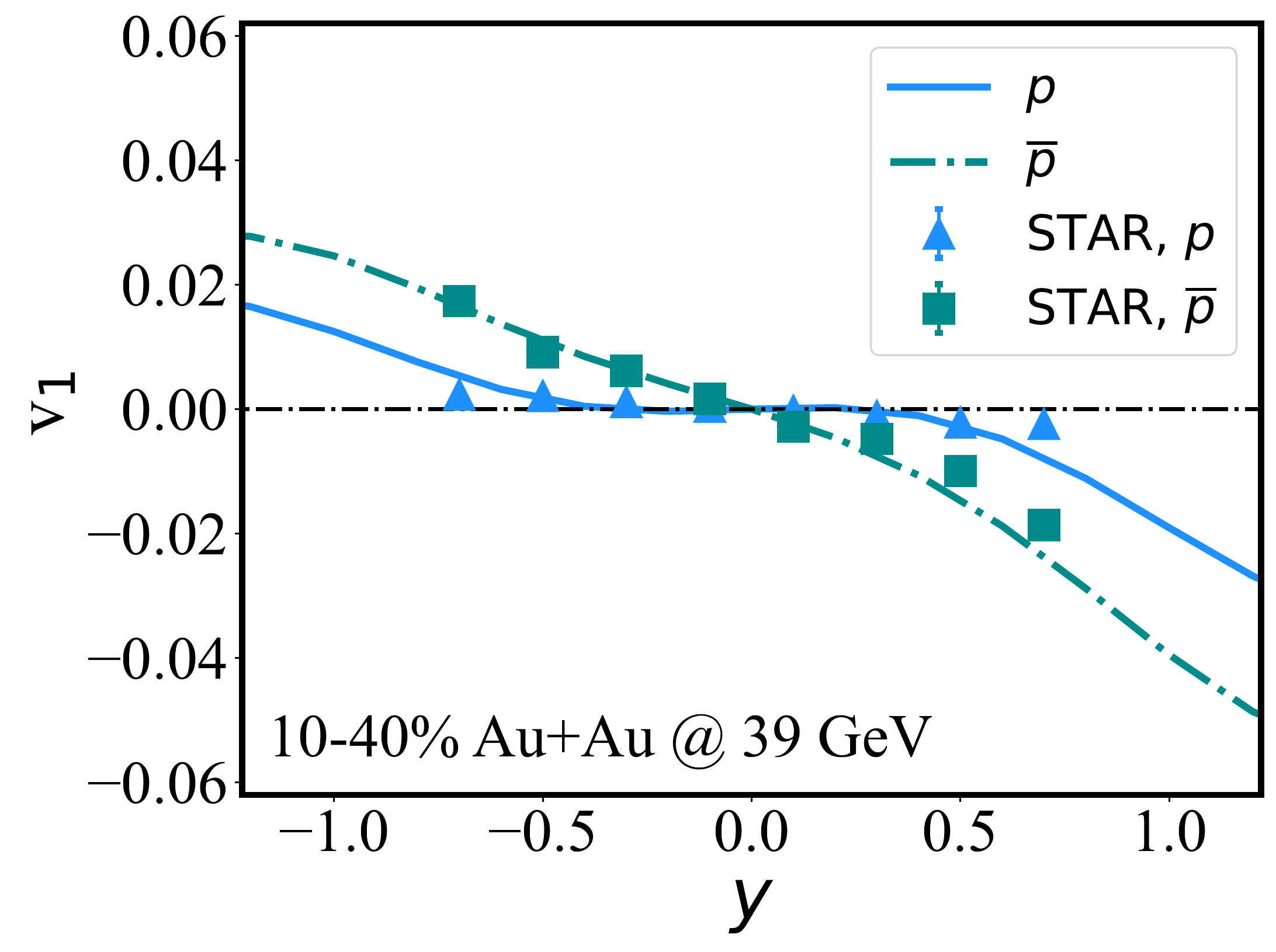}
\includegraphics[width=0.32\textwidth]{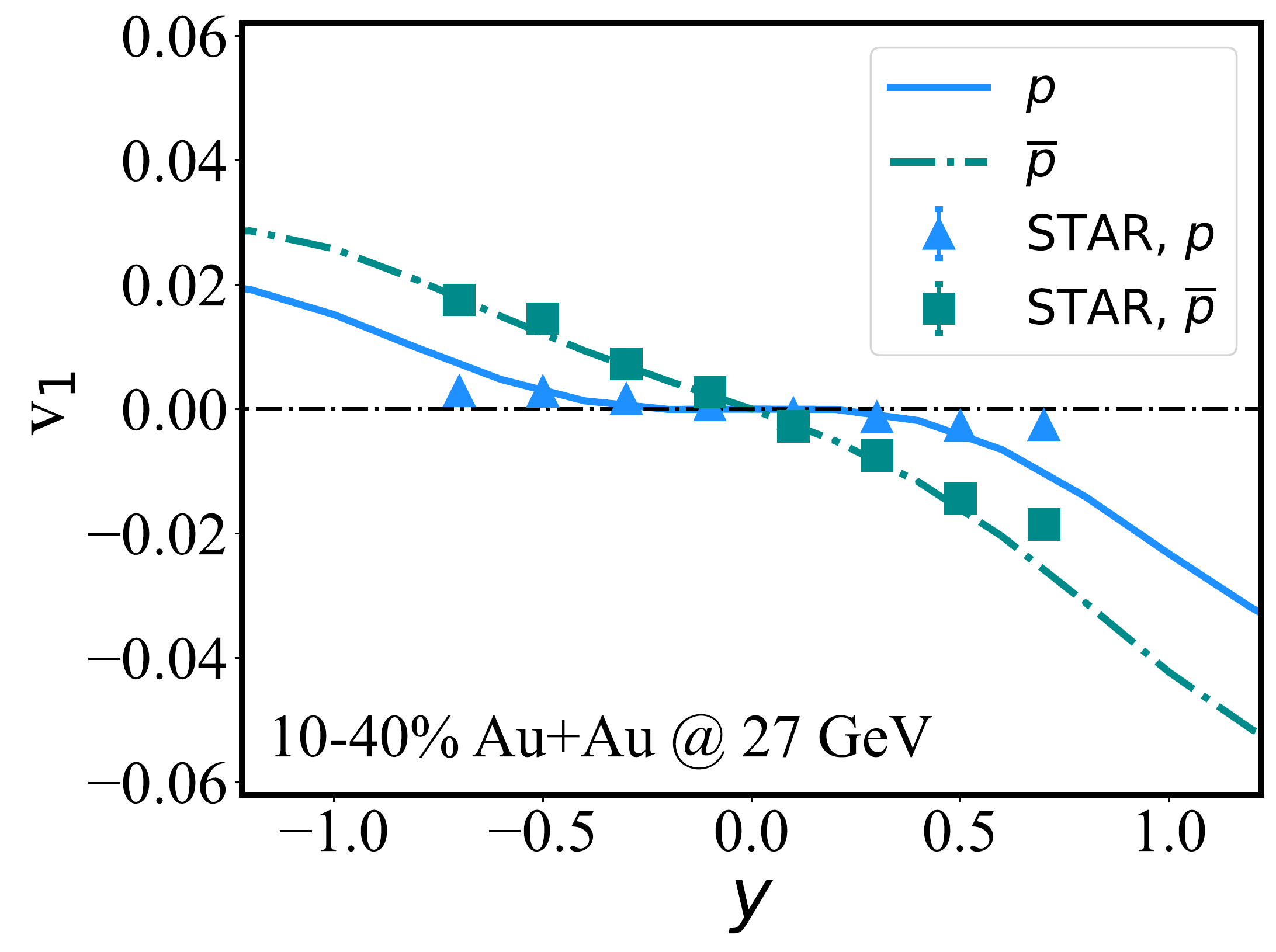}
\vspace{-4pt}
\includegraphics[width=0.32\textwidth]{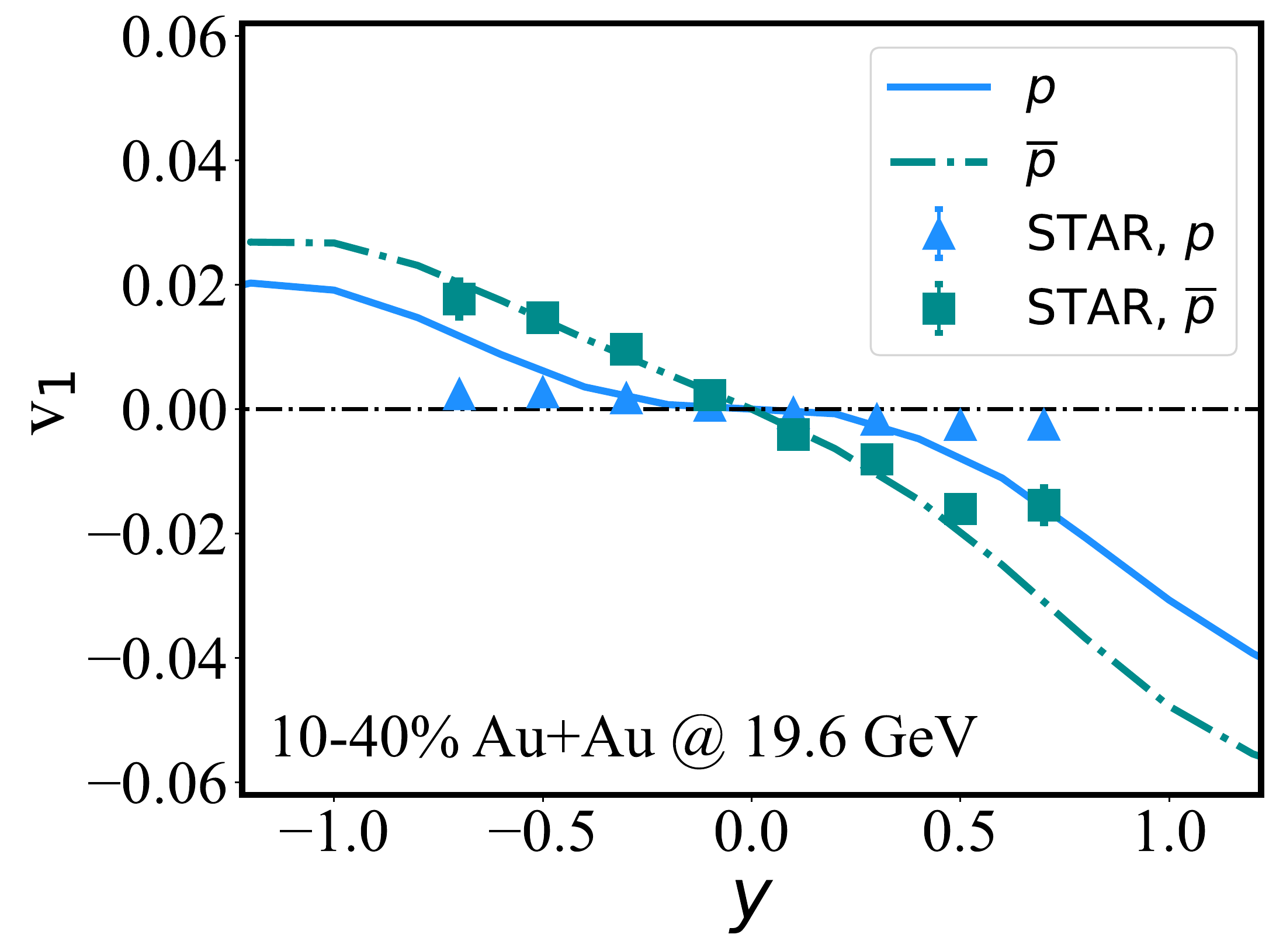}
\includegraphics[width=0.32\textwidth]{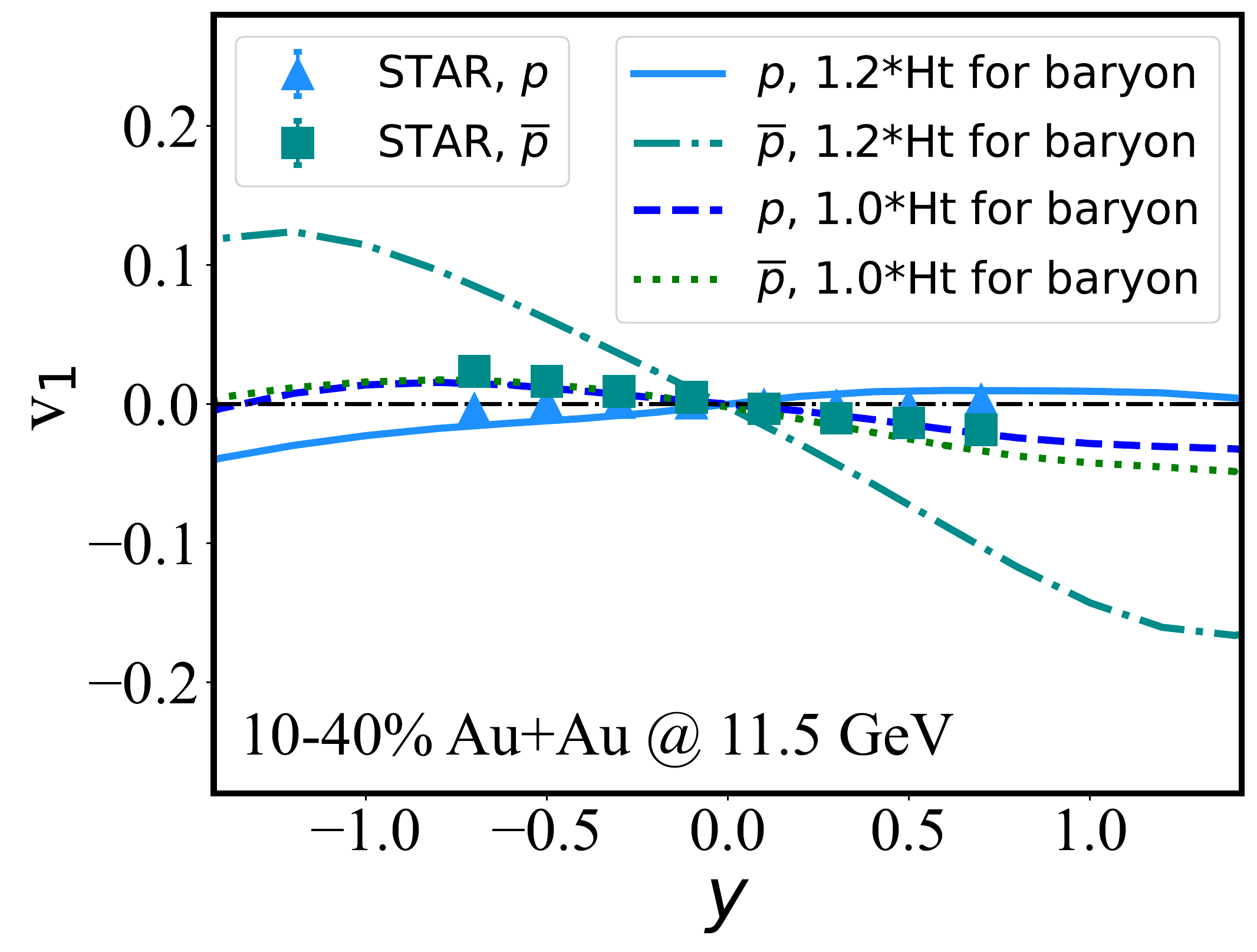}
\includegraphics[width=0.32\textwidth]{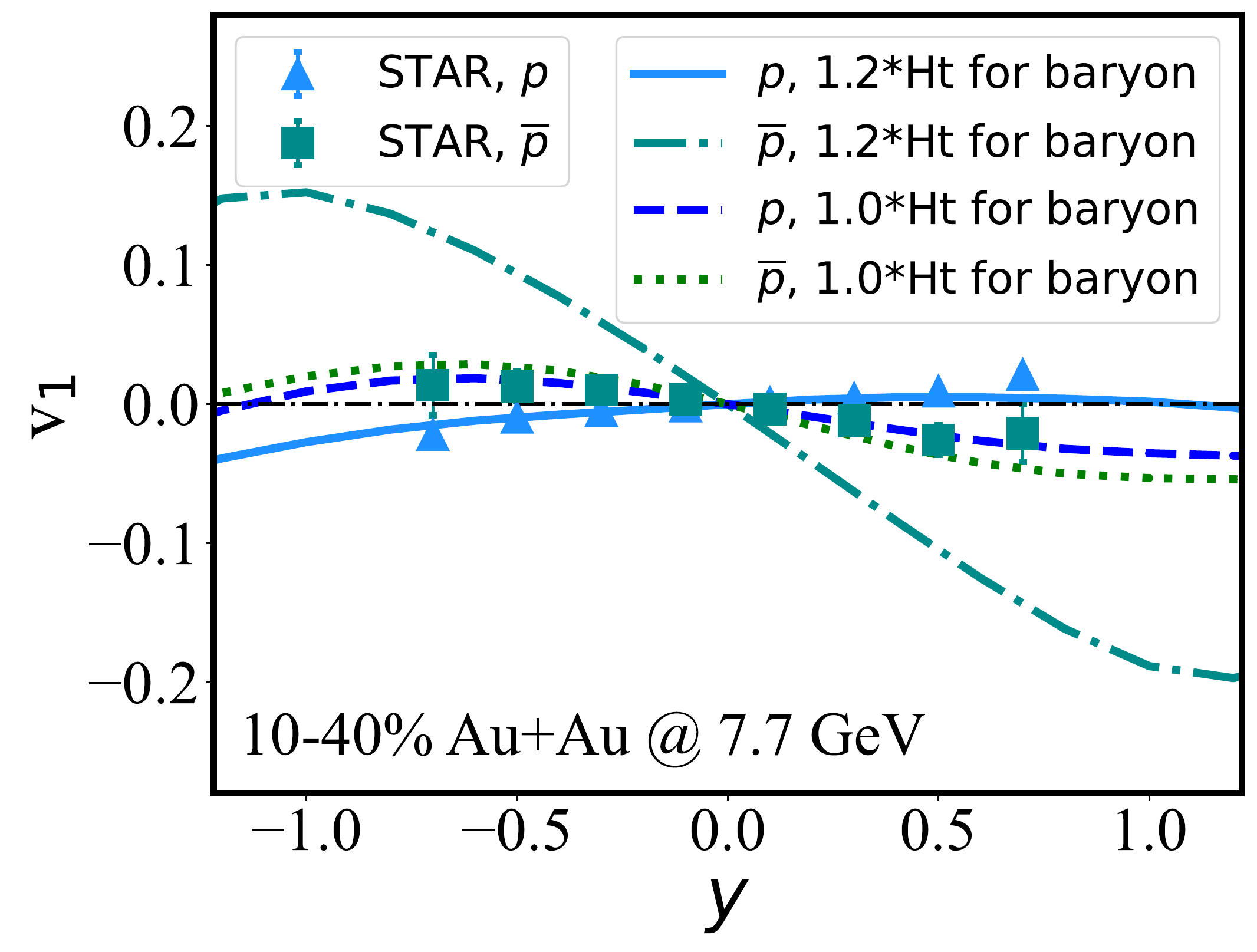}
\caption{(Color online) Rapidity dependence of the directed flow coefficient of protons and antiprotons in 10-40\% Au+Au collisions at the BES energies, compared between our model calculation and the STAR data~\cite{STAR:2014clz}.}
\label{f:protron-v1}
\end{figure*}

\subsection{Directed flow coefficients of $\pi^{+}$, $p$ and $\bar{p}$}
\label{sec:3-2}
In this subsection, we study the directed flow coefficients of $\pi^{+}$, $p$ and $\bar{p}$ in 10-40\% Au+Au collisions at the BES energies (7.7 - 62.4 GeV). As the first order Fourier component of the azimuthal angle distribution, the directed flow coefficient at a given rapidity can be calculated as
\begin{equation}
\begin{aligned}
v_{1}(y)=\langle\cos(\phi-\Psi_{1})\rangle=\frac{\int\cos(\phi-\Psi_{1})\frac{dN}{dy d\phi}d\phi}{\int\frac{dN}{dy d\phi}d\phi},
\label{eq:v1}
\end{aligned}
\end{equation}
where $\Psi_{1}$ is the first order event plane angle of a nucleus-nucleus collision.
Since we use a smooth initial condition for the energy density and baryon number density distributions,
event-by-event fluctuations are ignored in the present work.
Therefore, the event plane here is the same as the spectator plane characterized
by the deflected neutrons in experimental measurements.
Effects of the initial-state fluctuations on the final-state hadron $v_1$ will be left for our future exploration. 

Shown in Fig.~\ref{f:pionv1} is the directed flow of $\pi^{+}$ as a function of rapidity in 10-40\% Au+Au collisions at the BES energies. Here, $v_1$ is analyzed using soft hadrons within $0<p_\mathrm{T}<3.0$~GeV.
One can see that our calculation provides a reasonable description of the pion $v_{1}(y)$ around mid-rapidity ($y\in[-1,1]$) observed at STAR~\cite{STAR:2014clz}. In this work, the directed flow is contributed by two mechanisms, the longitudinally tilted geometry of the QGP medium [Eq.~(\ref{eq:mnccnu})], as discussed in Ref.~\cite{Jiang:2021ajc}, and the non-zero initial gradient of the longitudinal flow velocity along the direction of the impact parameter ($\partial v_z/\partial x$) [Eqs.~(\ref{eq:Ttautau}) -~(\ref{eq:ycm})], as proposed in Ref.~\cite{Shen:2020jwv, Ryu:2021lnx, Alzhrani:2022dpi}. 
Detailed discussions on how these two effects contribute to the hadron $v_1$ and how their corresponding model parameters ($H_\text{t}$ and $f_v$) are adjusted will be presented later in Sec.~\ref{sec:3-3}. In the last two panels for $\snn=11.5$ and 7.7~GeV, we also present calculations using a larger tilt parameter ($1.2 H_\text{t}$) for the initial baryon number density distribution. As expected, the pion $v_1$ is not sensitive to this baryon distribution. 


\begin{figure}[tbp!]
\begin{center}
\includegraphics[width=0.75\linewidth]{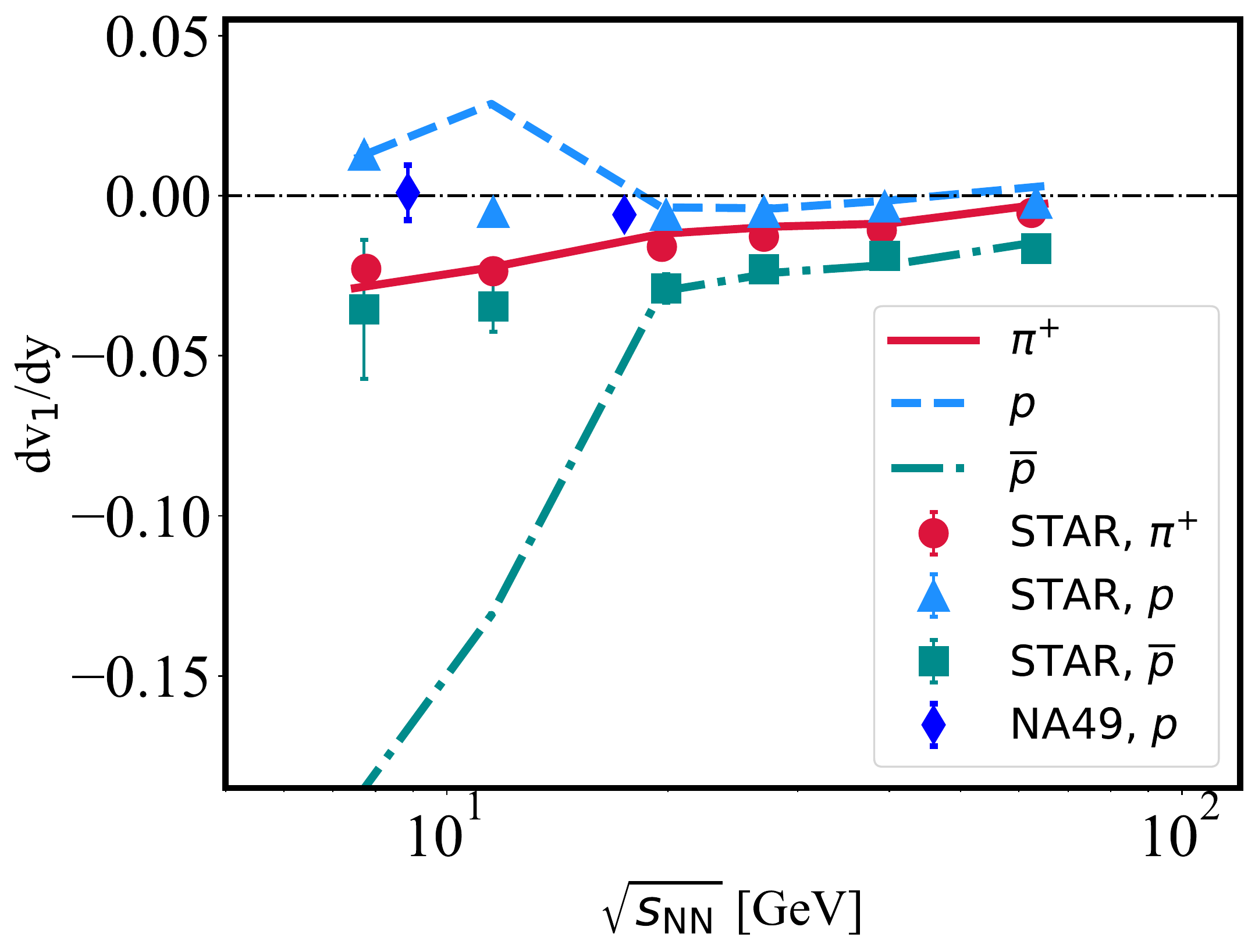}
\end{center}
\caption{(Color online) The slope of the rapidity dependence of the directed flow coefficients of pions, protons and antiprotons as functions of the colliding energy in 10-40\% Au+Au collisions. Results from our model calculation are compared to the STAR~\cite{STAR:2014clz} and NA49~\cite{NA49:2003njx} data. Here, $1.2 H_{\textrm{t}}$ is applied to the baryon number density distribution at $\snn=11.5$ and 7.7~GeV.}
\label{f:dv1dy}
\end{figure}

In Fig.~\ref{f:protron-v1}, we further study the rapidity dependence of $v_1$ for protons and antiprotons in 10-40\% Au+Au collisions at the BES energies. And for a more clear display of this rapidity dependence, its slope is extracted in Fig.~\ref{f:dv1dy} around mid rapidity ($y\in[-0.5,0.5]$) as a function of the colliding energy. Consistent with the findings of Ref.~\cite{Bozek:2022svy}, extending the tilted initial geometry from energy density to baryon number density leads to a separation of $v_1$ between protons and antiprotons. Within our model, using the same $H_\text{t}$ parameter is able to provide a good description of the experiment data when the colliding energy is not very low. However, at $\snn=11.5$ and 7.7~GeV, this minimal assumption seems insufficient. A slightly larger value of $H_\text{t}$ for baryon number density than for energy density is required to generate the increasing trend of proton $v_1$ with respect to $y$, as well as the splitting between protons and antiprotons observed at very low energies. This larger value of $H_\text{t}$ for the baryon density than the overall QGP medium might result from the possible phase transition in the pre-equilibrium stage~\cite{Ivanov:2014ioa}, the strong electromagnetic field~\cite{Rybicki:2013qla} and hadronic scatterings~\cite{Wu:2021fjf}.

As shown in Fig.~\ref{f:dv1dy}, with the increase of colliding energy, the splitting of $v_1$ between protons and antiprotons becomes smaller. This can be understood with the weaker tilt of the baryon number density distribution and the slower longitudinal flow velocity in higher energy collisions, and can be confirmed by the increasing values of $H_\text{t}$ and $f_v$ extracted in Tab.~\ref{table:parameters} as $\snn$ becomes smaller. Our calculation indicates the essential role of the geometric distribution of the baryon number density in producing the baryon and anti-baryon $v_1$. It also suggests the necessity of utilizing pions and protons (antiprotons) together to simultaneously constrain both the medium geometry and its longitudinal flow profile.

\subsection{Dependence of $dv_{1}/dy$ on the medium geometry and the longitudinal flow velocity}
\label{sec:3-3}

Both the tilted QGP profile and the longitudinal flow velocity gradient contribute to the directed flow of hadrons in heavy-ion collisions. In this subsection, we investigate how these two effects compete with each other and illustrate how the two parameters $H_\text{t}$ and $f_v$ are determined in our model calculation.

\begin{figure}[tbp!]
\begin{center}
\includegraphics[width=0.75\linewidth]{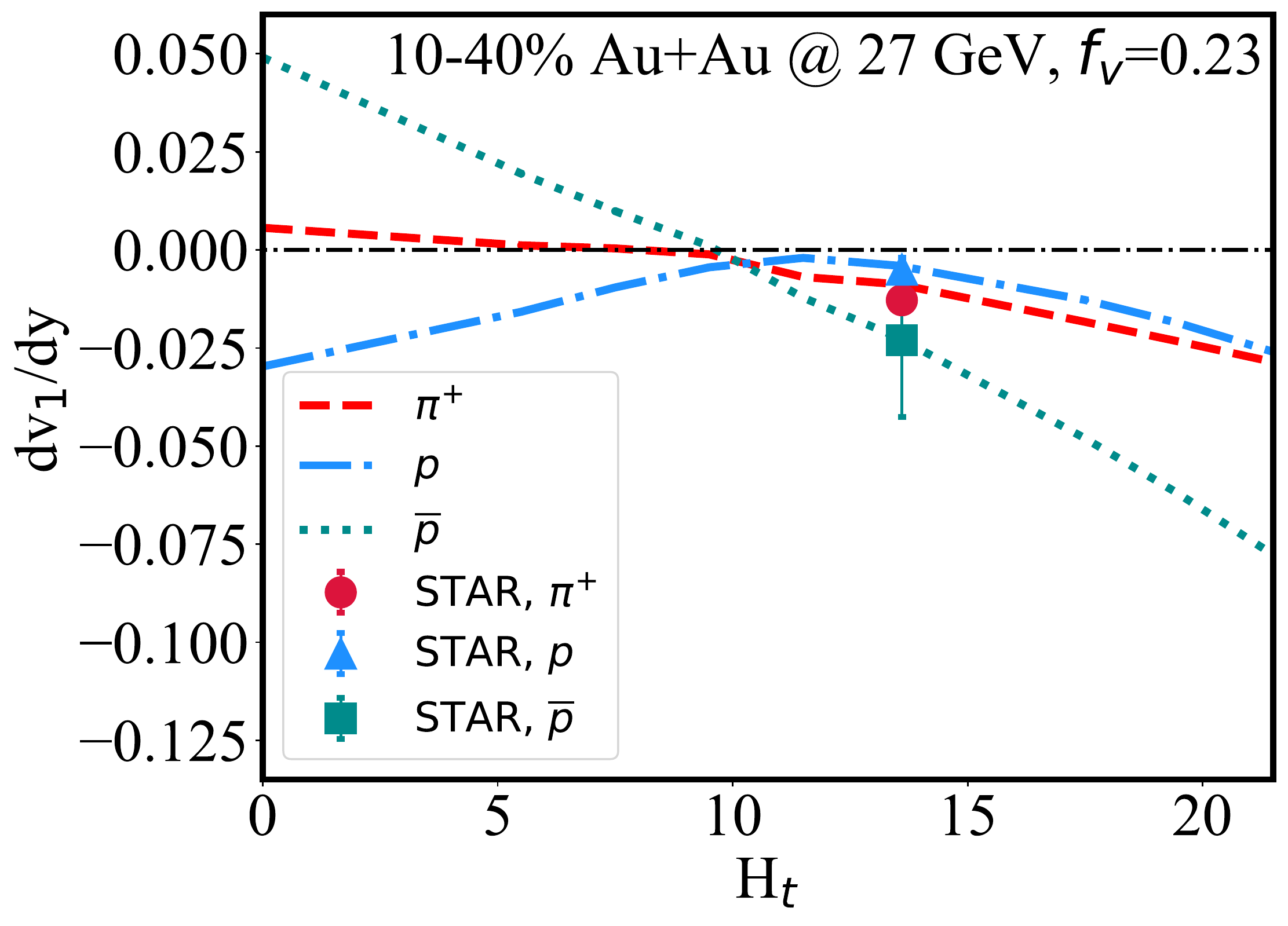}\\~~~
\includegraphics[width=0.75\linewidth]{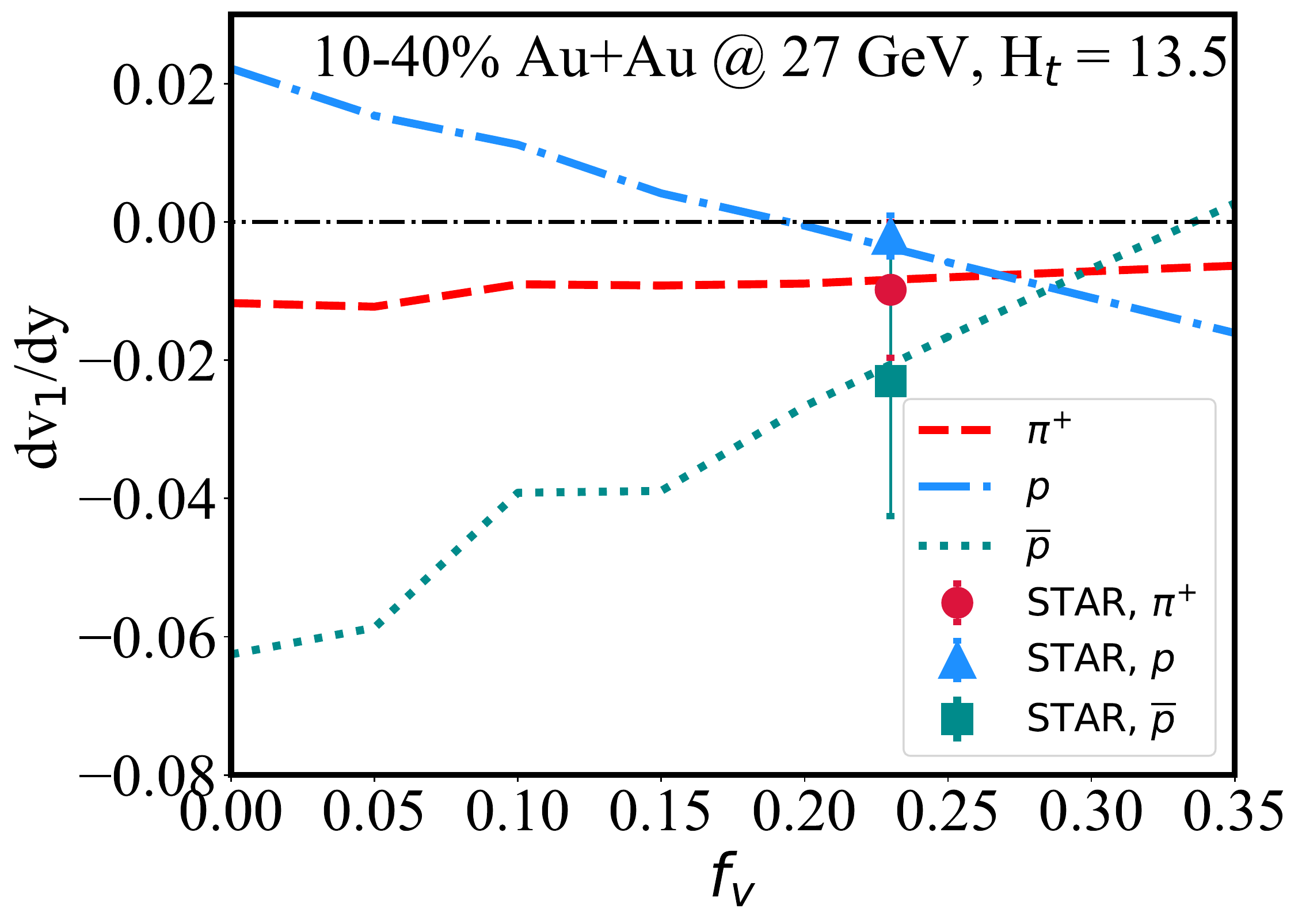}~
\end{center}
\caption{(Color online) Upper panel: the slope of the directed flow coefficient $dv_{1}/dy$ as a function of the tilt parameter $H_\text{t}$ when $f_v$ is fixed, in 10-40\% Au+Au collisions at $\snn=27$ GeV. Lower panel: the slope of the directed flow coefficient as a function of the longitudinal rapidity fraction parameter $f_{v}$ when $H_\text{t}$ is fixed. The experimental data are from the STAR Collaboration~\cite{STAR:2014clz}. }
\label{f:fv}
\end{figure}

In the upper panel of Fig.~\ref{f:fv}, we show the slope of the directed flow coefficient $dv_{1}/dy$ around mid-rapidity for $\pi^{+}$, proton and antiproton as a function of the $H_\text{t}$ parameter in 10-40\% Au+Au collisions at $\snn=$ 27 GeV, while $f_v$ is fixed at 0.23. When the longitudinal velocity profile is fixed, we observe the increase of $H_\text{t}$ from 0 to 23 leads to a decrease of the slope from positive to negative values for pions and antiprotons. In other words, with finite initial longitudinal flow velocity, the $v_1$ of pions and antiprotons are positive (negative) in the forward (backward) rapidity region when $H_\text{t}$ is small, but flip when $H_\text{t}$ is large. Since pions and antiprotons are mainly composed of partons newly produced by nuclear collisions, their $v_1$ both follow the energy density distribution of the QGP. On the other hand, protons carry baryons deposited by the beam nuclei and therefore their $v_1$ is significantly affected by the distribution of the initial baryon number density and show different features compared to pions and antiprotons.

In the lower panel of Fig.~\ref{f:fv}, we show the slope of $v_{1}(y)$ as a function $f_v$ when $H_\text{t}$ is fixed at 13.5. Compared to Fig.~\ref{f:fv}, we observe different dependences of $dv_{1}/dy$ on the medium deformation and its longitudinal flow velocity. While the slopes of pions and antiprotons decrease as the medium becomes more tilted (or $H_\text{t}$ increases), the opposite is seen when the longitudinal flow velocity (or $f_v$) increases. 

Figure~\ref{f:fv} suggests the sensitivity of the slope of $v_1(y)$ to both the medium geometry and its longitudinal flow profile in the initial state. Here, $H_\text{t}=13.5$ and $f_v=0.23$ provide the best simultaneous description of the $v_1$ of pions, protons and antiprotons at $\snn=27$~GeV. Values of these two parameters at other collision energies are constrained in the same way in our work.

\subsection{Global polarization}
\label{sec:3-4}

Apart from the directed flow, the longitudinal flow velocity gradient can also induce polarization of constituent partons inside the QGP via the spin-orbit coupling~\cite{Liang:2004ph,Liang:2004xn,Huang:2011ru,Ryu:2021lnx,Alzhrani:2022dpi,STAR:2017ckg,Li:2021zwq,Guo:2019joy,Wu:2022mkr}. It has also been found in Ref.~\cite{Li:2022pyw} that the global polarization can be affected by the initial geometry of the QGP medium when the same initial longitudinal velocity field is applied. Therefore, due to their similar origins, directed flow and global polarization should be correlated with each other. This correlation has recently been investigated in Refs.~\cite{Ivanov:2020wak,Ryu:2021lnx}. In this subsection, using the same model setup as previously implemented for studying the hadron $v_1$, we further explore the global polarization of $\Lambda$ and $\overline{\Lambda}$ in heavy-ion collisions. 


We assume quarks have reached local thermal equilibrium on their freezeout hypersurface. 
Meanwhile, the spin of quarks or hadrons are not modified during particlization and resonance decay~\cite{Wu:2021fjf,Yi:2021ryh,Yi:2021unq,Wu:2022mkr}.
Then, the polarization pseudo vector for spin 1/2 fermions can be obtained using the modified Cooper-Frye formalism as~\citep{Becattini:2013fla,Fang:2016vpj},
\begin{equation}
\mathcal{S}^{\mu}(\mathbf{p})=\frac{\int d \Sigma \cdot p \mathcal{J}_{5}^{\mu}(p, X)}{2 m \int d \Sigma \cdot \mathcal{N}(p, X)},
\end{equation}
where $\mathcal{J}^{\mu}_5$ is the axial charge current density and $\mathcal{N}^{\mu}(p, X)$ is the number density of fermions in the phase space.
Following the quantum kinetic theory \citep{Yi:2021ryh,Hidaka:2017auj,Yi:2021unq},
$\mathcal{S}^{\mu}(\mathbf{p})$ can be decomposed into different sources,
\begin{eqnarray}
\mathcal{S}^{\mu}(\mathbf{p}) & = & \mathcal{S}_{\textrm{thermal}}^{\mu}(\mathbf{p})
+\mathcal{S}_{\textrm{shear}}^{\mu}(\mathbf{p})+\mathcal{S}_{\textrm{accT}}^{\mu}(\mathbf{p})   \nonumber \\
& &+\mathcal{S}_{\textrm{chemical}}^{\mu}(\mathbf{p})+\mathcal{S}_{\textrm{EB}}^{\mu}(\mathbf{p}),
\end{eqnarray}
where
\begin{eqnarray}
\mathcal{S}_{\textrm{thermal}}^{\mu}(\mathbf{p}) & = & \int d\Sigma^{\sigma}F_{\sigma}\epsilon^{\mu\nu\alpha\beta}p_{\nu}\partial_{\alpha}\frac{u_{\beta}}{T},\nonumber \\
\mathcal{S}_{\textrm{shear}}^{\mu}(\mathbf{p}) & = & \int d\Sigma^{\sigma}F_{\sigma}  \frac{\epsilon^{\mu\nu\alpha\beta}p_{\nu} u_{\beta}}{(u\cdot p)T}
 \nonumber \\
 & &\times  p^{\rho}(\partial_{\rho}u_{\alpha}+\partial_{\alpha}u_{\rho}-u_{\rho}Du_{\alpha}), \nonumber \\
\mathcal{S}_{\textrm{accT}}^{\mu}(\mathbf{p}) & = & -\int d\Sigma^{\sigma}F_{\sigma}\frac{\epsilon^{\mu\nu\alpha\beta}p_{\nu}u_{\alpha}}{T}
\left(Du_{\beta}-\frac{\partial_{\beta}T}{T}\right),\nonumber \\
\mathcal{S}_{\textrm{chemical}}^{\mu}(\mathbf{p}) & = & 2\int d\Sigma^{\sigma}F_{\sigma}\frac{1}{(u\cdot p)}\epsilon^{\mu\nu\alpha\beta}p_{\alpha}u_{\beta}\partial_{\nu}\frac{\mu}{T},\nonumber \\
\mathcal{S}_{\textrm{EB}}^{\mu}(\mathbf{p}) & = & 2\int d\Sigma^{\sigma}F_{\sigma}\left[\frac{\epsilon^{\mu\nu\alpha\beta}p_{\alpha}u_{\beta}E_{\nu}}{(u\cdot p)T}+\frac{B^{\mu}}{T}\right],\nonumber \label{eq:S_all}  \\
\label{eq:5S}
\end{eqnarray}
with
\begin{align}
&F^{\mu} = \frac{\hbar}{8m_{\Lambda}\Phi(\mathbf{p})}p^{\mu}f_\text{eq}(1-f_\text{eq}), \nonumber \\
&\Phi(\mathbf{p}) = \int d\Sigma^{\mu}p_{\mu}f_\text{eq}.
\label{eq:def_N}
\end{align}
The five terms in Eq.~(\ref{eq:5S}) represent polarization induced by the thermal vorticity ($\mathcal{S}_{\textrm{thermal}}^{\mu}$), the shear tensor ($\mathcal{S}_{\textrm{shear}}^{\mu}$),
the fluid acceleration minus temperature gradient ($\mathcal{S}_{\textrm{accT}}^{\mu}$),
the gradient of chemical potential over temperature ($\mathcal{S}_{\textrm{chemical}}^{\mu}$),
and the external electromagnetic field ($\mathcal{S}_{\textrm{EB}}^{\mu}$), respectively.
Detailed expressions of these terms can be found in Refs.~\cite{Yi:2021ryh,Hidaka:2017auj,Yi:2021unq,Wu:2021fjf} or derived from the statistic model~\cite{Becattini:2021suc,Becattini:2021iol} and the Kubo formula~\cite{Liu:2020dxg,Liu:2021uhn,Fu:2021pok,Fu:2022myl}.
Here, $S^\mu_{\textrm{shear}}$ and $S^\mu_{\textrm{chemical}}$ are also named as the shear-induced polarization (SIP) and the baryonic spin Hall effect (SHE) in literature.
Since the electromagnetic field decay rapidly in heavy-ion collisions, the $\mathcal{S}_{\textrm{EB}}^{\mu}$ term is neglected in our current work.

The average polarization vector in the rest frame of $\Lambda$ (or $\bar{\Lambda}$) is then given by
\begin{eqnarray}
\vec{P}^{*}(\mathbf{p}) = \vec{P}(\mathbf{p})-\frac{\vec{P}(\mathbf{p}) \cdot \vec{p}}{p^{0}(p^{0}+m)}\vec{p},
\end{eqnarray}
where 
\begin{eqnarray}
P^{\mu}(\mathbf{p}) \equiv \frac{1}{s} \mathcal{S}^{\mu}(\mathbf{p}),
\end{eqnarray}
with $s=1/2$ being the spin of the particle.
After averaging over the transverse momentum, one obtains the local polarization as
\begin{eqnarray}
\langle \vec{P}(\phi_p) \rangle = \frac{\int_{y_{\text{min}}}^{y_{\text{max}}}dy \int_{p_\text{T\text{min}}}^{p_\text{T\text{max}}}p_\text{T}dp_\text{T}
[ \Phi (\mathbf{p})\vec{P}^{*}(\mathbf{p})]}{\int_{y_{\text{min}}}^{y_{\text{max}}}dy \int_{p_\text{T\text{min}}}^{p_\text{T\text{max}}}p_\text{T}dp_\text{T} \Phi(\mathbf{p}) },
\label{eq:localP}
\end{eqnarray}
in which $\phi_p$ is the azimuthal angle, and $\Phi(\mathbf{p})$ is an integration on the freezeout hypersurface defined in Eq.~(\ref{eq:def_N}). 
In the present study, the mass of $\Lambda$ (or $\overline{\Lambda}$) is set as $m = 1.116$~GeV, and the kinematic regions for analyzing the hyperon polarization are $p_\text{T} \in [0.5~\text{GeV},~3.0~\text{GeV}]$ and $y\in [-1,~1]$. Finally, the global polarization of $\Lambda$ and $\overline{\Lambda}$ is obtained by further averaging $\vec{P}^{*}(\mathbf{p})$ over $\phi_p$ in Eq.~(\ref{eq:localP}).


\begin{figure}[tbp!]
\begin{center}
\includegraphics[width=0.75\linewidth]{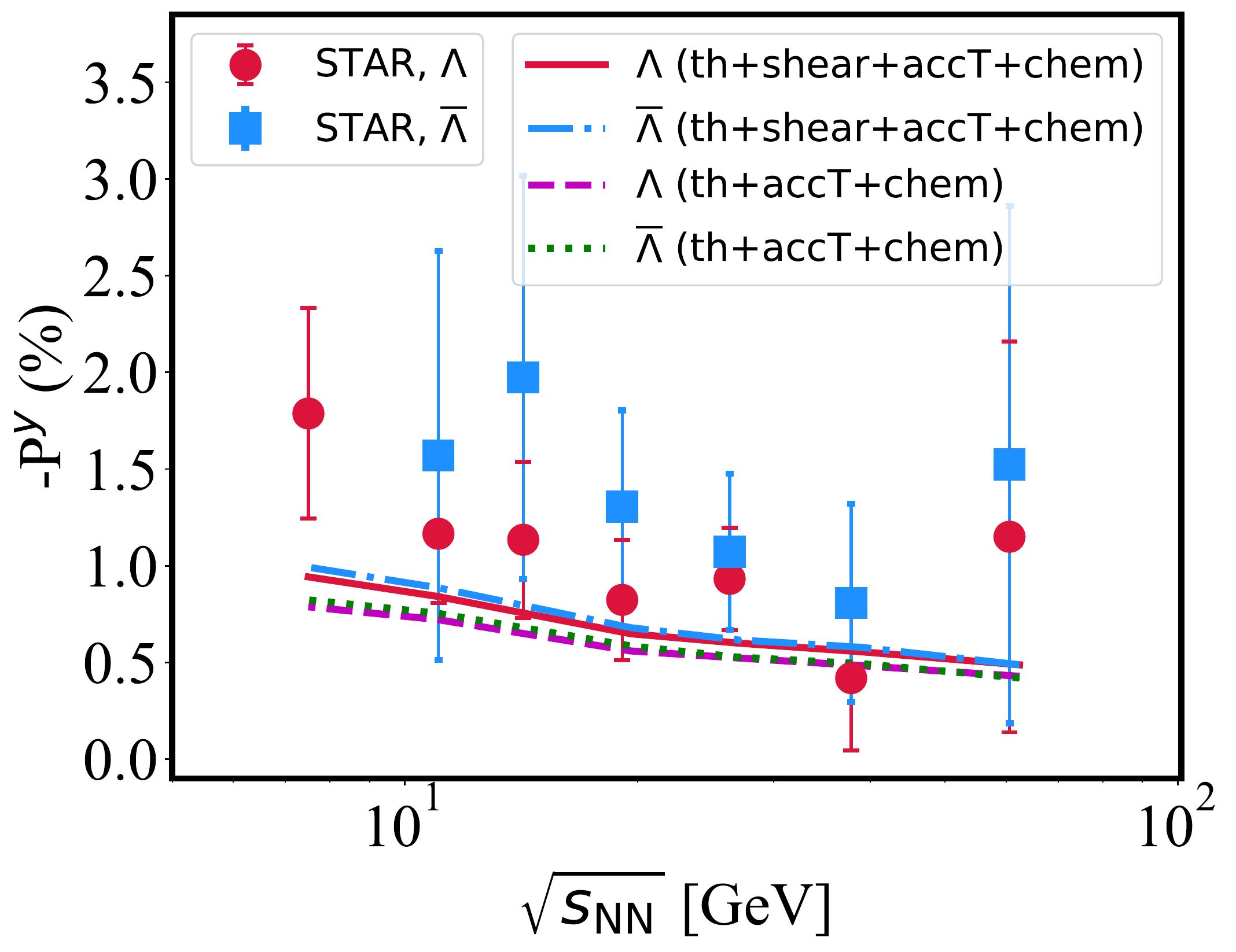}~\\
\includegraphics[width=0.75\linewidth]{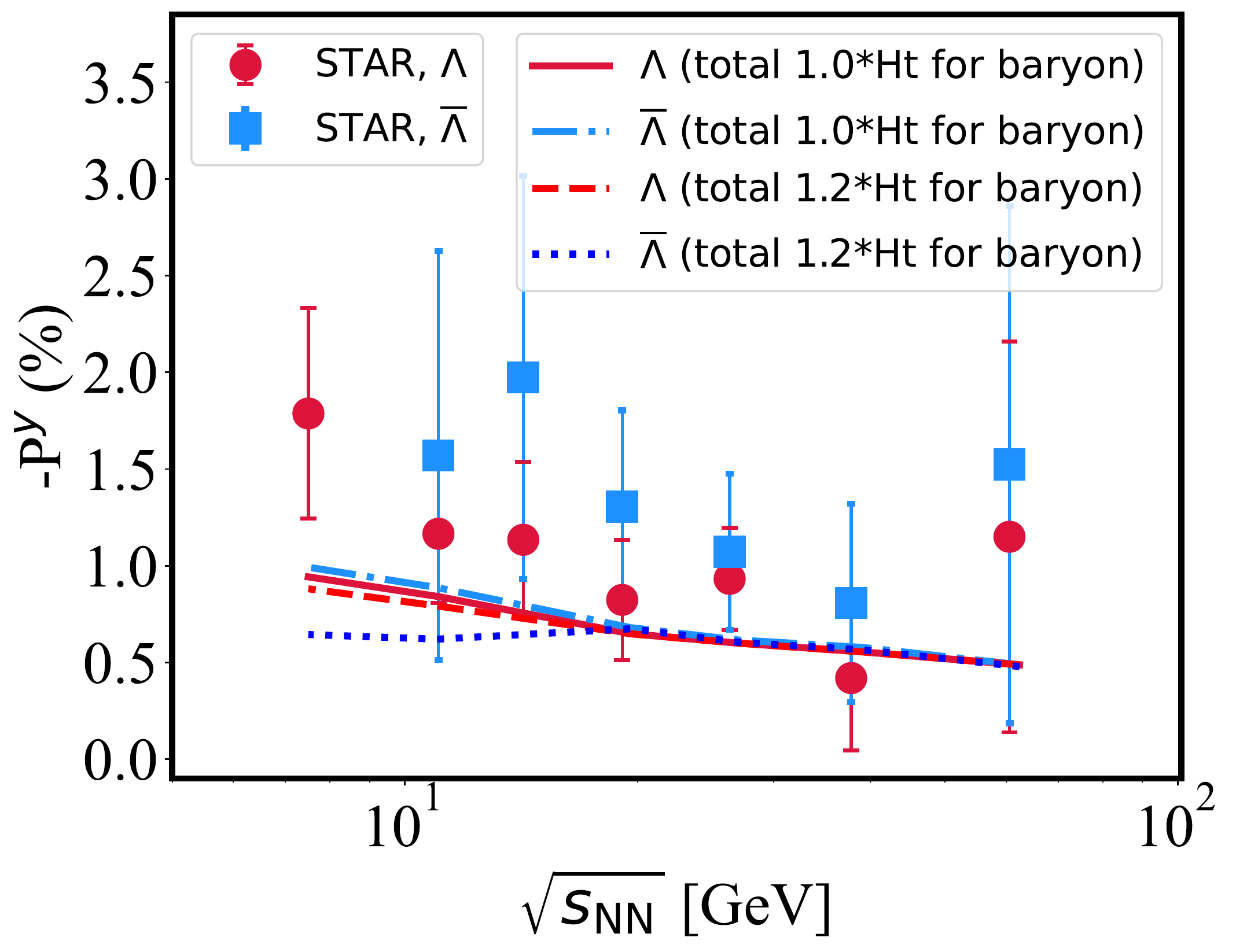}~
\end{center}
\caption{(Color online) The global polarization of $\Lambda$ and $\overline{\Lambda}$ as a function of the collision energy in 20-50\% Au+Au collisions, analyzed within $p_\text{T} \in [0.5~\text{GeV},3.0~\text{GeV}]$, $y\in [-1,1] $ and compared to the STAR data~\citep{STAR:2021beb} (rescaled by 0.877 due to the updated hyperon decay parameter~\cite{Ryu:2021lnx,ParticleDataGroup:2020ssz}). In the lower panel, ``total'' denotes ``thermal + shear + accT + chemical" for short.}
\label{f:pol}
\end{figure}

Shown in Fig.~\ref{f:pol} is the global polarization of $\Lambda$ and $\bar{\Lambda}$ hyperons along the out-of-plane direction as a function of the collision energy in 20-50\% Au+Au collisions. Using the model parameters $H_\text{t}$ and $f_{v}$ extracted from the directed flow of pions, protons and antiprotons as discussed earlier, our model calculation also provides a reasonable description of the hyperon polarization here. This helps further validate our modeling of the tilted geometry of the QGP together with its longitudinal flow gradient, and also confirms the correlation between the directed flow and the global polarization. 
We have confirmed that the total amount of polarization presented here is mainly contributed by the thermal vorticity term. The shear tensor relies on the tilted geometry of the medium: a counterclockwise tilt in the reaction plane induces a negative shear tensor, while a clockwise tilt induces a positive shear tensor. Therefore, as shown in the upper panel of Fig.~\ref{f:pol}, introducing the shear contribution increases the magnitude of $-P_y$. In the lower panel of Fig.~\ref{f:pol}, we study the effects of the possible larger value of $H_\mathrm{t}$ for the net baryon number density than for the energy density at low energies ($\snn=11.5$ and 7.7~GeV), as previously suggested by the proton $v_1$. After introducing this stronger tilt of baryon number density distribution, the global polarization becomes smaller because of a positive contribution from the chemical term to $P_y$. We note that the difference between $\Lambda$ and $\bar{\Lambda}$ polarization relies on both the geometry of baryon distribution and the longitudinal flow velocity. This is why our current result appears different from the earlier study~\citep{Wu:2022mkr} using the same hydrodynamic calculation but different initial condition. The electromagnetic field may cause further difference between their polarization, which has not been included in our current study. 


\section{Conclusions}
\label{v1section4}

We have developed an initial condition model for studying the directed flow and global polarization of identified hadrons in heavy-ion collisions across the BES energies. 
The Glauber model has been extended such that effects of the tilted geometry of both the energy density and the net baryon number density distribution have been included. The initial longitudinal flow velocity profile has also been introduced.
By combining this initial condition with a hydrodynamic simulation of the QGP evolution, we have provided a satisfactory description of the transverse momentum spectra of identified particles ($\pi^+$, $K^+$, $p$ and $\bar{p}$) from $\snn=7.7$ to 62.4~GeV. The directed flow coefficient of these identified hadrons, especially the splitting of $v_1$ between protons and antiprotons have been investigated in detail.


Our calculation shows the essential role of the tilted medium geometry and the early-time longitudinal flow velocity in generating the directed flow of hadrons. While both the strength of tilt (or the $H_\text{t}$ parameter) and the fractional longitudinal momentum deposition (or $f_v$) become larger at lower collisional energies, they have different impacts on the slope of $v_1(y)$ of different hadron species. An increasing $H_\text{t}$ results in a decrease of $dv_1/dy$ around mid-rapidity for both pions and antiprotons, while an increasing $f_v$ causes their increase. The opposite trend could be seen for protons, which is strongly affected by the baryon number density distribution deposited by the colliding beam. Therefore, a simultaneous comparison of the pion, proton and antiproton $v_1$ to their experimental data sets a tight constraint on the inhomogeneous distribution of the initial energy and baryon number density, and the longitudinal flow velocity profile of the nuclear matter created in non-central heavy-ion collisions. The initial geometry and flow velocity extracted from the hadron $v_1$ is further tested with the global polarization of $\Lambda$ and $\bar{\Lambda}$ hyperons that is also affected by the medium geometry and the fluid velocity gradient in the longitudinal direction. Using the same hydrodynamic framework, we obtain a reasonable description of the hyperon polarization across the BES energies, and find that the separation of global polarization between $\Lambda$ and $\bar{\Lambda}$ could also be sensitive to the tilted geometry of the net baryon number density distribution. Note that the splitting of  $v_1$ between protons and antiprotons is not reflected in a large splitting of the polarization between $\Lambda$ and $\bar{\Lambda}$ from $\snn=7.7$ to 62.4~GeV, implying different sensitivities of these two observables to the baryon number density distribution.

%

Our study constitutes a step forward in understanding the origin of the splitting of $v_1$ between different particle species produced in Au+Au collisions at $\snn=7.7$ - 62.4~GeV. While the $v_1$ of protons and antiprotons is described well for energies $\snn = 19.6$~GeV and above, additional effects besides the deformed medium geometry and the initial longitudinal flow velocity gradient should be explored for the splitting of $v_1$ between protons and antiprotons at lower energies.
For example, the electromagnetic field produced in non-central heavy-ion collisions results in directional drift of charged quarks ($u,d,s$) and therefore different values of $v_1$ between different final state charged particles. Additionally, the light hadron $v_1$ can also be affected by the decelerated baryon flow and hadronic cascade after the QGP expansion, especially at lower collision energy~\cite{Shen:2020jwv,Ryu:2021lnx,Bozek:2022svy}. Therefore, a combination of hydrodynamic model and afterburner hadronic transport is necessary for a better constraint on the initial state. Furthermore, due to the limit of our current smooth initial condition, our calculation is restricted to the rapidity odd component of $v_1$. 
The rapidity even component, especially its non-trivial $p_\mathrm{T}$ dependence even at mid-rapidity~\cite{Teaney:2010vd,Luzum:2010fb,Gale:2012rq}is another interesting topic to investigate after our initialization method is extended to include event-by-event fluctuations. Last but not least, the initial condition and the QGP profile we propose here can be straightforwardly coupled to studies of hard probes in low energy collisions, such as the collective flow of high $p_\text{T}$ hadrons and heavy quarks~\cite{Jiang:2022vxe} and their correlations~\cite{Ma:2017ybx,ALICE:2019oyn}. 
These will be addressed in our upcoming efforts.

\begin{acknowledgements}
This work was supported by the National Natural Science Foundation of China (NSFC) under Grant Nos.~11935007, 12175122 and 2021-867, Guangdong Major Project of Basic and Applied Basic Research No.~2020B0301030008, the Natural Science Foundation of Hubei Province No.~2021CFB272, the Education Department of Hubei Province of China with Young Talents Project No.~Q20212703, the Open Foundation of Key Laboratory of Quark and Lepton Physics (MOE) No.~QLPL202104 and the Xiaogan Natural Science Foundation under Grant No.~XGKJ2021010016. Xiang-Yu Wu would like to thank the support from the CCNU Youbo Program.
\end{acknowledgements}

\bibliographystyle{unsrt}
\bibliography{clv3_v1ref}

\begin{thebibliography}{100}

\bibitem{Ollitrault:1992bk}
Jean-Yves Ollitrault.
\newblock {Anisotropy as a signature of transverse collective flow}.
\newblock {\em Phys. Rev. D}, 46:229--245, 1992.

\bibitem{Rischke:1995ir}
Dirk~H. Rischke, S.~Bernard, and J.~A. Maruhn.
\newblock {Relativistic hydrodynamics for heavy ion collisions. 1. General
  aspects and expansion into vacuum}.
\newblock {\em Nucl. Phys. A}, 595:346--382, 1995.

\bibitem{Sorge:1996pc}
H.~Sorge.
\newblock {Elliptical flow: A Signature for early pressure in ultrarelativistic
  nucleus-nucleus collisions}.
\newblock {\em Phys. Rev. Lett.}, 78:2309--2312, 1997.

\bibitem{Bass:1998vz}
S.A. Bass, M.~Gyulassy, H.~Stoecker, and W.~Greiner.
\newblock {Signatures of quark gluon plasma formation in high-energy heavy ion
  collisions: A Critical review}.
\newblock {\em J. Phys. G}, 25:R1--R57, 1999.

\bibitem{Aguiar:2001ac}
C.~E. Aguiar, Y.~Hama, T.~Kodama, and T.~Osada.
\newblock {Event-by-event fluctuations in hydrodynamical description of heavy
  ion collisions}.
\newblock {\em Nucl. Phys. A}, 698:639--642, 2002.

\bibitem{Shuryak:2003xe}
E.~Shuryak.
\newblock {Why does the quark gluon plasma at RHIC behave as a nearly ideal
  fluid?}
\newblock {\em Prog. Part. Nucl. Phys.}, 53:273--303, 2004.

\bibitem{Heinz:2013th}
Ulrich Heinz and Raimond Snellings.
\newblock {Collective flow and viscosity in relativistic heavy-ion collisions}.
\newblock {\em Ann. Rev. Nucl. Part. Sci.}, 63:123--151, 2013.

\bibitem{Busza:2018rrf}
Wit Busza, Krishna Rajagopal, and Wilke van~der Schee.
\newblock {Heavy Ion Collisions: The Big Picture, and the Big Questions}.
\newblock {\em Ann. Rev. Nucl. Part. Sci.}, 68:339--376, 2018.

\bibitem{Aoki:2006we}
Y.~Aoki, G.~Endrodi, Z.~Fodor, S.~D. Katz, and K.~K. Szabo.
\newblock {The Order of the quantum chromodynamics transition predicted by the
  standard model of particle physics}.
\newblock {\em Nature}, 443:675--678, 2006.

\bibitem{Friman:2011zz}
Bengt Friman, Claudia Hohne, Jorn Knoll, Stefan Leupold, Jorgen Randrup, Ralf
  Rapp, and Peter Senger.
\newblock {The CBM physics book: Compressed baryonic matter in laboratory
  experiments}.
\newblock {\em Lect. Notes Phys.}, 814, 2011.

\bibitem{Rischke:1995pe}
Dirk~H. Rischke, Yaris P\"urs\"un, Joachim~A. Maruhn, Horst Stoecker, and
  Walter Greiner.
\newblock {The Phase transition to the quark - gluon plasma and its effects on
  hydrodynamic flow}.
\newblock {\em Acta Phys. Hung. A}, 1:309--322, 1995.

\bibitem{Odyniec:2015iaa}
Grazyna Odyniec.
\newblock {Future of the beam energy scan program at RHIC}.
\newblock {\em EPJ Web Conf.}, 95:03027, 2015.

\bibitem{Kekelidze:2016wkp}
V.~Kekelidze, A.~Kovalenko, R.~Lednicky, V.~Matveev, I.~Meshkov, A.~Sorin, and
  G.~Trubnikov.
\newblock {Prospects for the dense baryonic matter research at NICA}.
\newblock {\em Nucl. Phys. A}, 956:846--849, 2016.

\bibitem{J-PARCHeavy-Ion:2016ikk}
H.~Sako et~al.
\newblock {Studies of high density baryon matter with high intensity heavy-ion
  beams at J-PARC}.
\newblock {\em Nucl. Phys. A}, 956:850--853, 2016.

\bibitem{E877:1997zjw}
J.~Barrette et~al.
\newblock {Proton and pion production relative to the reaction plane in Au + Au
  collisions at AGS energies}.
\newblock {\em Phys. Rev. C}, 56:3254--3264, 1997.

\bibitem{E895:2000maf}
H.~Liu et~al.
\newblock {Sideward flow in Au + Au collisions between 2-A-GeV and 8-A-GeV}.
\newblock {\em Phys. Rev. Lett.}, 84:5488--5492, 2000.

\bibitem{CBM:2016kpk}
T.~Ablyazimov et~al.
\newblock {Challenges in QCD matter physics --The scientific programme of the
  Compressed Baryonic Matter experiment at FAIR}.
\newblock {\em Eur. Phys. J. A}, 53(3):60, 2017.

\bibitem{STAR:2021iop}
Mohamed Abdallah et~al.
\newblock {Cumulants and correlation functions of net-proton, proton, and
  antiproton multiplicity distributions in Au+Au collisions at energies
  available at the BNL Relativistic Heavy Ion Collider}.
\newblock {\em Phys. Rev. C}, 104(2):024902, 2021.

\bibitem{STAR:2014clz}
L.~Adamczyk et~al.
\newblock {Beam-Energy Dependence of the Directed Flow of Protons, Antiprotons,
  and Pions in Au+Au Collisions}.
\newblock {\em Phys. Rev. Lett.}, 112(16):162301, 2014.

\bibitem{STAR:2017okv}
L.~Adamczyk et~al.
\newblock {Beam-Energy Dependence of Directed Flow of $\Lambda$,
  $\bar{\Lambda}$, $K^\pm$, $K^0_s$ and $\phi$ in Au+Au Collisions}.
\newblock {\em Phys. Rev. Lett.}, 120(6):062301, 2018.

\bibitem{STAR:2021ozh}
M.~S. Abdallah et~al.
\newblock {Light nuclei collectivity from~$\sqrt{s_{NN}}$ = 3 GeV Au+Au
  collisions at RHIC}.
\newblock {\em Phys. Lett. B}, 827:136941, 2022.

\bibitem{Luo:2020pef}
Xiaofeng Luo, Shusu Shi, Nu~Xu, and Yifei Zhang.
\newblock {A Study of the Properties of the QCD Phase Diagram in High-Energy
  Nuclear Collisions}.
\newblock {\em Particles}, 3(2):278--307, 2020.

\bibitem{Bzdak:2019pkr}
Adam Bzdak, Shinichi Esumi, Volker Koch, Jinfeng Liao, Mikhail Stephanov, and
  Nu~Xu.
\newblock {Mapping the Phases of Quantum Chromodynamics with Beam Energy Scan}.
\newblock {\em Phys. Rept.}, 853:1--87, 2020.

\bibitem{Sun:2020zxy}
Kai-Jia Sun, Feng Li, and Che~Ming Ko.
\newblock {Effects of QCD critical point on light nuclei production}.
\newblock {\em Phys. Lett. B}, 816:136258, 2021.

\bibitem{STAR:2017ieb}
L.~Adamczyk et~al.
\newblock {Beam Energy Dependence of Jet-Quenching Effects in Au+Au Collisions
  at $\sqrt{s_{_{ \mathrm{NN}}}}$ = 7.7, 11.5, 14.5, 19.6, 27, 39, and 62.4
  GeV}.
\newblock {\em Phys. Rev. Lett.}, 121(3):032301, 2018.

\bibitem{Wu:2022vbu}
Jing Wu, Shanshan Cao, and Feng Li.
\newblock {Partonic Critical Opalescence and Its Impact on the Jet Quenching
  Parameter $\hat{q}$}.
\newblock {\em arXiv:2208.14297}.

\bibitem{Voloshin:1994mz}
S.~Voloshin and Y.~Zhang.
\newblock {Flow study in relativistic nuclear collisions by Fourier expansion
  of Azimuthal particle distributions}.
\newblock {\em Z. Phys. C}, 70:665--672, 1996.

\bibitem{Bilandzic:2010jr}
A.~Bilandzic, R.~Snellings, and S.~Voloshin.
\newblock {Flow analysis with cumulants: Direct calculations}.
\newblock {\em Phys. Rev. C}, 83:044913, 2011.

\bibitem{STAR:2004jwm}
J.~Adams et~al.
\newblock {Azimuthal anisotropy in Au+Au collisions at s(NN)**(1/2) = 200-GeV}.
\newblock {\em Phys. Rev. C}, 72:014904, 2005.

\bibitem{STAR:2019clv}
J.~Adam et~al.
\newblock {First Observation of the Directed Flow of $D^{0}$ and
  $\overline{D^0}$ in Au+Au Collisions at $\sqrt{s_{\rm NN}}$ = 200 GeV}.
\newblock {\em Phys. Rev. Lett.}, 123(16):162301, 2019.

\bibitem{ALICE:2019sgg}
S.~Acharya et~al.
\newblock {Probing the effects of strong electromagnetic fields with
  charge-dependent directed flow in Pb-Pb collisions at the LHC}.
\newblock {\em Phys. Rev. Lett.}, 125(2):022301, 2020.

\bibitem{STAR:2019vcp}
Jaroslav Adam et~al.
\newblock {Bulk properties of the system formed in $Au+Au$ collisions at
  $\sqrt{s_{\mathrm{NN}}}$ =14.5 GeV at the BNL STAR detector}.
\newblock {\em Phys. Rev. C}, 101(2):024905, 2020.

\bibitem{Gyulassy:1981nq}
M.~Gyulassy, K.~A. Frankel, and Horst Stoecker.
\newblock {DO NUCLEI FLOW AT HIGH-ENERGIES?}
\newblock {\em Phys. Lett. B}, 110:185--188, 1982.

\bibitem{Gustafsson:1984ka}
H.~A. Gustafsson et~al.
\newblock {Collective Flow Observed in Relativistic Nuclear Collisions}.
\newblock {\em Phys. Rev. Lett.}, 52:1590--1593, 1984.

\bibitem{Lisa:2000ip}
Michael~Annan Lisa, Ulrich~W. Heinz, and Urs~Achim Wiedemann.
\newblock {Tilted pion sources from azimuthally sensitive HBT interferometry}.
\newblock {\em Phys. Lett. B}, 489:287--292, 2000.

\bibitem{PHENIX:2003qra}
S.~S. Adler et~al.
\newblock {Elliptic flow of identified hadrons in Au+Au collisions at
  s(NN)**(1/2) = 200-GeV}.
\newblock {\em Phys. Rev. Lett.}, 91:182301, 2003.

\bibitem{ALICE:2010suc}
K~Aamodt et~al.
\newblock {Elliptic flow of charged particles in Pb-Pb collisions at 2.76 TeV}.
\newblock {\em Phys. Rev. Lett.}, 105:252302, 2010.

\bibitem{CMS:2012zex}
S.~Chatrchyan et~al.
\newblock {Measurement of the elliptic anisotropy of charged particles produced
  in PbPb collisions at $\sqrt{s}_{NN}$=2.76 TeV}.
\newblock {\em Phys. Rev. C}, 87(1):014902, 2013.

\bibitem{Nara:2016phs}
Yasushi Nara, Harri Niemi, Akira Ohnishi, and Horst St\"ocker.
\newblock {Examination of directed flow as a signature of the softest point of
  the equation of state in QCD matter}.
\newblock {\em Phys. Rev. C}, 94(3):034906, 2016.

\bibitem{Chatterjee:2017ahy}
S.~Chatterjee and P.~Bo\.zek.
\newblock {Large directed flow of open charm mesons probes the three
  dimensional distribution of matter in heavy ion collisions}.
\newblock {\em Phys. Rev. Lett.}, 120(19):192301, 2018.

\bibitem{Singha:2016mna}
S.~Singha, P.~Shanmuganathan, and D.~Keane.
\newblock {The first moment of azimuthal anisotropy in nuclear collisions from
  AGS to LHC energies}.
\newblock {\em Adv. High Energy Phys.}, 2016:2836989, 2016.

\bibitem{Zhang:2018wlk}
Chao Zhang, Jiamin Chen, Xiaofeng Luo, Feng Liu, and Y.~Nara.
\newblock {Beam energy dependence of the squeeze-out effect on the directed and
  elliptic flow in Au + Au collisions in the high baryon density region}.
\newblock {\em Phys. Rev. C}, 97(6):064913, 2018.

\bibitem{Guo:2017mkf}
Chong-Qiang Guo, Chun-Jian Zhang, and Jun Xu.
\newblock {Revisiting directed flow in relativistic heavy-ion collisions from a
  multiphase transport model}.
\newblock {\em Eur. Phys. J. A}, 53(12):233, 2017.

\bibitem{Parida:2022lmt}
Tribhuban Parida and Sandeep Chatterjee.
\newblock {Splitting of elliptic flow in a tilted fireball}.
\newblock {\em Phys. Rev. C}, 106(4):044907, 2022.

\bibitem{Bozek:2022svy}
Piotr Bozek.
\newblock {Splitting of proton-antiproton directed flow in relativistic
  heavy-ion collisions}.
\newblock {\em Phys. Rev. C}, 106(6):L061901, 2022.

\bibitem{Bozek:2011ua}
Piotr Bozek.
\newblock {Flow and interferometry in 3+1 dimensional viscous hydrodynamics}.
\newblock {\em Phys. Rev. C}, 85:034901, 2012.

\bibitem{Jiang:2021ajc}
Ze-Fang Jiang, Shanshan Cao, Xiang-Yu Wu, C.~B. Yang, and Ben-Wei Zhang.
\newblock {Longitudinal distribution of initial energy density and directed
  flow of charged particles in relativistic heavy-ion collisions}.
\newblock {\em Phys. Rev. C}, 105(3):034901, 2022.

\bibitem{Jiang:2021foj}
Ze-Fang Jiang, C.~B. Yang, and Qi~Peng.
\newblock {Directed flow of charged particles within idealized viscous
  hydrodynamics at energies available at the BNL Relativistic Heavy Ion
  Collider and at the CERN Large Hadron Collider}.
\newblock {\em Phys. Rev. C}, 104(6):064903, 2021.

\bibitem{Shen:2020jwv}
Chun Shen and S.~Alzhrani.
\newblock {Collision-geometry-based 3D initial condition for relativistic
  heavy-ion collisions}.
\newblock {\em Phys. Rev. C}, 102(1):014909, 2020.

\bibitem{Ryu:2021lnx}
Sangwook Ryu, Vahidin Jupic, and Chun Shen.
\newblock {Probing early-time longitudinal dynamics with the
  \ensuremath{\Lambda} hyperon's spin polarization in relativistic heavy-ion
  collisions}.
\newblock {\em Phys. Rev. C}, 104(5):054908, 2021.

\bibitem{NA49:2003njx}
C.~Alt et~al.
\newblock {Directed and elliptic flow of charged pions and protons in Pb + Pb
  collisions at 40-A-GeV and 158-A-GeV}.
\newblock {\em Phys. Rev. C}, 68:034903, 2003.

\bibitem{Chen:2009xc}
J.~Y. Chen, J.~X. Zuo, X.~Z. Cai, F.~Liu, Y.~G. Ma, and A.~H. Tang.
\newblock {Energy Dependence of Directed Flow in Au+Au Collisions from a
  Multi-phase Transport Model}.
\newblock {\em Phys. Rev. C}, 81:014904, 2010.

\bibitem{Steinheimer:2014pfa}
J.~Steinheimer, J.~Auvinen, H.~Petersen, M.~Bleicher, and H.~St\"ocker.
\newblock {Examination of directed flow as a signal for a phase transition in
  relativistic nuclear collisions}.
\newblock {\em Phys. Rev. C}, 89(5):054913, 2014.

\bibitem{Konchakovski:2014gda}
V.~P. Konchakovski, W.~Cassing, Yu.~B. Ivanov, and V.~D. Toneev.
\newblock {Examination of the directed flow puzzle in heavy-ion collisions}.
\newblock {\em Phys. Rev. C}, 90(1):014903, 2014.

\bibitem{Guo:2012qi}
Yao Guo, Feng Liu, and Aihong Tang.
\newblock {Directed flow of transported and non-transported protons in Au+Au
  collisions from UrQMD model}.
\newblock {\em Phys. Rev. C}, 86:044901, 2012.

\bibitem{Ivanov:2014ioa}
Yu.~B. Ivanov and A.~A. Soldatov.
\newblock {Directed flow indicates a cross-over deconfinement transition in
  relativistic nuclear collisions}.
\newblock {\em Phys. Rev. C}, 91(2):024915, 2015.

\bibitem{Parida:2022zse}
Tribhuban Parida and Sandeep Chatterjee.
\newblock {Directed flow of light flavor hadrons for Au+Au collisions at
  $\sqrt{S_{NN}}=$ 7.7-200 GeV}.
\newblock {\em arXiv: 2211.15659}.

\bibitem{Parida:2022ppj}
Tribhuban Parida and Sandeep Chatterjee.
\newblock {Directed flow in a baryonic fireball}.
\newblock {\em arXiv: 2211.15729}.

\bibitem{Du:2022yok}
Lipei Du, Chun Shen, Sangyong Jeon, and Charles Gale.
\newblock {Probing initial baryon stopping and equation of state with
  rapidity-dependent directed flow of identified particles}.
\newblock {\em arXiv: 2211.16408}.

\bibitem{Pang:2016igs}
Long-Gang Pang, H.~Petersen, Qun Wang, and Xin-Nian Wang.
\newblock {Vortical Fluid and $\Lambda$ Spin Correlations in High-Energy
  Heavy-Ion Collisions}.
\newblock {\em Phys. Rev. Lett.}, 117(19):192301, 2016.

\bibitem{Pang:2018zzo}
Long-Gang Pang, H.~Petersen, and Xin-Nian Wang.
\newblock {Pseudorapidity distribution and decorrelation of anisotropic flow
  within the open-computing-language implementation CLVisc hydrodynamics}.
\newblock {\em Phys. Rev. C}, 97(6):064918, 2018.

\bibitem{Wu:2018cpc}
Xiang-Yu Wu, Long-Gang Pang, Guang-You Qin, and Xin-Nian Wang.
\newblock {Longitudinal fluctuations and decorrelations of anisotropic flows at
  energies available at the CERN Large Hadron Collider and at the BNL
  Relativistic Heavy Ion Collider}.
\newblock {\em Phys. Rev. C}, 98(2):024913, 2018.

\bibitem{Wu:2021fjf}
Xiang-Yu Wu, Guang-You Qin, Long-Gang Pang, and Xin-Nian Wang.
\newblock {(3+1)-D viscous hydrodynamics at finite net baryon density:
  Identified particle spectra, anisotropic flows, and flow fluctuations across
  energies relevant to the beam-energy scan at RHIC}.
\newblock {\em Phys. Rev. C}, 105(3):034909, 2022.

\bibitem{Bozek:2010bi}
P.~Bozek and I.~Wyskiel.
\newblock {Directed flow in ultrarelativistic heavy-ion collisions}.
\newblock {\em Phys. Rev. C}, 81:054902, 2010.

\bibitem{Loizides:2017ack}
C.~Loizides, J.~Kamin, and D.~d'Enterria.
\newblock {Improved Monte Carlo Glauber predictions at present and future
  nuclear colliders}.
\newblock {\em Phys. Rev. C}, 97(5):054910, 2018.
\newblock [Erratum: Phys.Rev.C 99, 019901 (2019)].

\bibitem{STAR:2021mii}
Mohamed Abdallah et~al.
\newblock {Search for the chiral magnetic effect with isobar collisions at
  $\sqrt {s_{NN}}$=200 GeV by the STAR Collaboration at the BNL Relativistic
  Heavy Ion Collider}.
\newblock {\em Phys. Rev. C}, 105(1):014901, 2022.

\bibitem{Jiang:2022uoe}
Ze-Fang Jiang, Shanshan Cao, Wen-Jing Xing, Xiang-Yu Wu, C.~B. Yang, and
  Ben-Wei Zhang.
\newblock {Probing the initial longitudinal density profile and electromagnetic
  field in ultrarelativistic heavy-ion collisions with heavy quarks}.
\newblock {\em Phys. Rev. C}, 105(5):054907, 2022.

\bibitem{Shen:2017bsr}
Chun Shen and Bj\"orn Schenke.
\newblock {Dynamical initial state model for relativistic heavy-ion
  collisions}.
\newblock {\em Phys. Rev. C}, 97(2):024907, 2018.

\bibitem{Bialas:2004kt}
A.~Bialas and M.~Jezabek.
\newblock {Bremsstrahlung from color charges as a source of soft particle
  production in hadronic collisions}.
\newblock {\em Phys. Lett. B}, 590:233--238, 2004.

\bibitem{Li:2022pyw}
Xiaowen Li, Ze-Fang Jiang, Shanshan Cao, and Jian Deng.
\newblock {Evolution of global polarization in relativistic heavy-ion
  collisions within a perturbative approach}.
\newblock {\em Eur. Phys. J. C}, 83(1):96, 2023.

\bibitem{Alzhrani:2022dpi}
Sahr Alzhrani, Sangwook Ryu, and Chun Shen.
\newblock {$\Lambda$ spin polarization in event-by-event relativistic heavy-ion
  collisions}.
\newblock {\em Phys. Rev. C}, 106:014905, 2022.

\bibitem{Rybicki:2013qla}
Andrzej Rybicki and Antoni Szczurek.
\newblock {Spectator induced electromagnetic effect on directed flow in heavy
  ion collisions}.
\newblock {\em Phys. Rev. C}, 87(5):054909, 2013.

\bibitem{Jiang:2020big}
Ze~Fang Jiang, Duan She, C.B. Yang, and Defu Hou.
\newblock {Perturbation solutions of relativistic viscous hydrodynamics
  forlongitudinally expanding fireballs}.
\newblock {\em Chin. Phys. C}, 44(8):084107, 2020.

\bibitem{Jiang:2018qxd}
Ze~Fang Jiang, C.B. Yang, Chi Ding, and Xiang-Yu Wu.
\newblock {Pseudo-rapidity distribution from a perturbative solution of viscous
  hydrodynamics for heavy ion collisions at RHIC and LHC}.
\newblock {\em Chin. Phys. C}, 42(12):123103, 2018.

\bibitem{Denicol:2012cn}
G.S. Denicol, H.~Niemi, E.~Molnar, and D.H. Rischke.
\newblock {Derivation of transient relativistic fluid dynamics from the
  Boltzmann equation}.
\newblock {\em Phys. Rev. D}, 85:114047, 2012.
\newblock [Erratum: Phys.Rev.D 91, 039902 (2015)].

\bibitem{Romatschke:2009im}
P.~Romatschke.
\newblock {New Developments in Relativistic Viscous Hydrodynamics}.
\newblock {\em Int. J. Mod. Phys. E}, 19:1--53, 2010.

\bibitem{Romatschke:2017ejr}
P.~Romatschke and U.~Romatschke.
\newblock {\em {Relativistic Fluid Dynamics In and Out of Equilibrium}}.
\newblock Cambridge Monographs on Mathematical Physics. Cambridge University
  Press, 5 2019.

\bibitem{Akamatsu:2018olk}
Yukinao Akamatsu, Masayuki Asakawa, Tetsufumi Hirano, Masakiyo Kitazawa, Kenji
  Morita, Koichi Murase, Yasushi Nara, Chiho Nonaka, and Akira Ohnishi.
\newblock {Dynamically integrated transport approach for heavy-ion collisions
  at high baryon density}.
\newblock {\em Phys. Rev. C}, 98(2):024909, 2018.

\bibitem{Denicol:2018wdp}
Gabriel~S. Denicol, Charles Gale, Sangyong Jeon, Akihiko Monnai, Bj\"orn
  Schenke, and Chun Shen.
\newblock {Net baryon diffusion in fluid dynamic simulations of relativistic
  heavy-ion collisions}.
\newblock {\em Phys. Rev. C}, 98(3):034916, 2018.

\bibitem{Monnai:2019hkn}
Akihiko Monnai, Bj\"orn Schenke, and Chun Shen.
\newblock {Equation of state at finite densities for QCD matter in nuclear
  collisions}.
\newblock {\em Phys. Rev. C}, 100(2):024907, 2019.

\bibitem{Monnai:2021kgu}
Akihiko Monnai, Bj\"orn Schenke, and Chun Shen.
\newblock {QCD Equation of State at Finite Chemical Potentials for Relativistic
  Nuclear Collisions}.
\newblock {\em Int. J. Mod. Phys. A}, 36(07):2130007, 2021.

\bibitem{McNelis:2021acu}
M.~McNelis and U.~Heinz.
\newblock {Modified equilibrium distributions for Cooper--Frye particlization}.
\newblock {\em Phys. Rev. C}, 103(6):064903, 2021.

\bibitem{Liang:2004ph}
Zuo-Tang Liang and Xin-Nian Wang.
\newblock {Globally polarized quark-gluon plasma in non-central A+A
  collisions}.
\newblock {\em Phys. Rev. Lett.}, 94:102301, 2005.
\newblock [Erratum: Phys.Rev.Lett. 96, 039901 (2006)].

\bibitem{Liang:2004xn}
Zuo-Tang Liang and Xin-Nian Wang.
\newblock {Spin alignment of vector mesons in non-central A+A collisions}.
\newblock {\em Phys. Lett. B}, 629:20--26, 2005.

\bibitem{Huang:2011ru}
Xu-Guang Huang, Pasi Huovinen, and Xin-Nian Wang.
\newblock {Quark Polarization in a Viscous Quark-Gluon Plasma}.
\newblock {\em Phys. Rev. C}, 84:054910, 2011.

\bibitem{STAR:2017ckg}
L.~Adamczyk et~al.
\newblock {Global $\Lambda$ hyperon polarization in nuclear collisions:
  evidence for the most vortical fluid}.
\newblock {\em Nature}, 548:62--65, 2017.

\bibitem{Li:2021zwq}
Hui Li, Xiao-Liang Xia, Xu-Guang Huang, and Huan~Zhong Huang.
\newblock {Global spin polarization of multistrange hyperons and feed-down
  effect in heavy-ion collisions}.
\newblock {\em Phys. Lett. B}, 827:136971, 2022.

\bibitem{Guo:2019joy}
Yu~Guo, Shuzhe Shi, Shengqin Feng, and Jinfeng Liao.
\newblock {Magnetic Field Induced Polarization Difference between Hyperons and
  Anti-hyperons}.
\newblock {\em Phys. Lett. B}, 798:134929, 2019.

\bibitem{Wu:2022mkr}
Xiang-Yu Wu, Cong Yi, Guang-You Qin, and Shi Pu.
\newblock {Local and global polarization of \ensuremath{\Lambda} hyperons
  across RHIC-BES energies: The roles of spin hall effect, initial condition,
  and baryon diffusion}.
\newblock {\em Phys. Rev. C}, 105(6):064909, 2022.

\bibitem{Ivanov:2020wak}
Yu.~B. Ivanov and A.~A. Soldatov.
\newblock {Correlation between global polarization, angular momentum, and flow
  in heavy-ion collisions}.
\newblock {\em Phys. Rev. C}, 102(2):024916, 2020.

\bibitem{Yi:2021ryh}
Cong Yi, Shi Pu, and Di-Lun Yang.
\newblock {Reexamination of local spin polarization beyond global equilibrium
  in relativistic heavy ion collisions}.
\newblock {\em Phys. Rev. C}, 104(6):064901, 2021.

\bibitem{Yi:2021unq}
Cong Yi, Shi Pu, Jian-Hua Gao, and Di-Lun Yang.
\newblock {Hydrodynamic helicity polarization in relativistic heavy ion
  collisions}.
\newblock {\em Phys. Rev. C}, 105(4):044911, 2022.

\bibitem{Becattini:2013fla}
F.~Becattini, V.~Chandra, L.~Del~Zanna, and E.~Grossi.
\newblock {Relativistic distribution function for particles with spin at local
  thermodynamical equilibrium}.
\newblock {\em Annals Phys.}, 338:32--49, 2013.

\bibitem{Fang:2016vpj}
Ren-hong Fang, Long-gang Pang, Qun Wang, and Xin-nian Wang.
\newblock {Polarization of massive fermions in a vortical fluid}.
\newblock {\em Phys. Rev. C}, 94(2):024904, 2016.

\bibitem{Hidaka:2017auj}
Yoshimasa Hidaka, Shi Pu, and Di-Lun Yang.
\newblock {Nonlinear Responses of Chiral Fluids from Kinetic Theory}.
\newblock {\em Phys. Rev. D}, 97(1):016004, 2018.

\bibitem{Becattini:2021suc}
F.~Becattini, M.~Buzzegoli, and A.~Palermo.
\newblock {Spin-thermal shear coupling in a relativistic fluid}.
\newblock {\em Phys. Lett. B}, 820:136519, 2021.

\bibitem{Becattini:2021iol}
F.~Becattini, M.~Buzzegoli, G.~Inghirami, I.~Karpenko, and A.~Palermo.
\newblock {Local Polarization and Isothermal Local Equilibrium in Relativistic
  Heavy Ion Collisions}.
\newblock {\em Phys. Rev. Lett.}, 127(27):272302, 2021.

\bibitem{Liu:2020dxg}
Shuai Y.~F. Liu and Yi~Yin.
\newblock {Spin Hall effect in heavy-ion collisions}.
\newblock {\em Phys. Rev. D}, 104(5):054043, 2021.

\bibitem{Liu:2021uhn}
Shuai Y.~F. Liu and Yi~Yin.
\newblock {Spin polarization induced by the hydrodynamic gradients}.
\newblock {\em JHEP}, 07:188, 2021.

\bibitem{Fu:2021pok}
Baochi Fu, Shuai Y.~F. Liu, Longgang Pang, Huichao Song, and Yi~Yin.
\newblock {Shear-Induced Spin Polarization in Heavy-Ion Collisions}.
\newblock {\em Phys. Rev. Lett.}, 127(14):142301, 2021.

\bibitem{Fu:2022myl}
Baochi Fu, Longgang Pang, Huichao Song, and Yi~Yin.
\newblock {Signatures of the spin Hall effect in hot and dense QCD matter}.
\newblock {\em arXiv:2201.12970}.

\bibitem{STAR:2021beb}
M.~S. Abdallah et~al.
\newblock {Global $\Lambda$-hyperon polarization in Au+Au collisions at $\sqrt
  {s_{NN}}$=3~GeV}.
\newblock {\em Phys. Rev. C}, 104(6):L061901, 2021.

\bibitem{ParticleDataGroup:2020ssz}
P.~A. Zyla et~al.
\newblock {Review of Particle Physics}.
\newblock {\em PTEP}, 2020(8):083C01, 2020.

\bibitem{Teaney:2010vd}
Derek Teaney and Li~Yan.
\newblock {Triangularity and Dipole Asymmetry in Heavy Ion Collisions}.
\newblock {\em Phys. Rev. C}, 83:064904, 2011.

\bibitem{Luzum:2010fb}
Matthew Luzum and Jean-Yves Ollitrault.
\newblock {Directed flow at midrapidity in heavy-ion collisions}.
\newblock {\em Phys. Rev. Lett.}, 106:102301, 2011.

\bibitem{Gale:2012rq}
Charles Gale, Sangyong Jeon, Bj\"orn Schenke, Prithwish Tribedy, and Raju
  Venugopalan.
\newblock {Event-by-event anisotropic flow in heavy-ion collisions from
  combined Yang-Mills and viscous fluid dynamics}.
\newblock {\em Phys. Rev. Lett.}, 110(1):012302, 2013.

\bibitem{Jiang:2022vxe}
Ze-Fang Jiang, Shanshan Cao, Wen-Jing Xing, Xiaowen Li, and Ben-Wei Zhang.
\newblock {Interactions between heavy quarks and tilted QGP fireballs in 200 A
  GeV Au+Au collisions}.
\newblock {\em Chin. Phys. C}, 47(2):024107, 2023.

\bibitem{Ma:2017ybx}
Long Ma, Xin Dong, Huan-Zhong Huang, and Yu-Gang Ma.
\newblock {Study of a background reconstruction method for the measurement of
  D-meson azimuthal angular correlations}.
\newblock {\em Nucl. Sci. Tech.}, 32(6):61, 2021.

\bibitem{ALICE:2019oyn}
Shreyasi Acharya et~al.
\newblock {Azimuthal correlations of prompt D mesons with charged particles in
  pp and p\textendash{}Pb collisions at $\sqrt{s_{NN}}$ = 5.02 TeV}.
\newblock {\em Eur. Phys. J. C}, 80(10):979, 2020.

\end{thebibliography}

\end{document}